\documentclass{scrartcl}
\pdfoutput=1

\usepackage[a4paper,top=2.5cm, bottom=3cm, left=2.5cm, right=2.5cm]{geometry}

\usepackage[
backend=biber,      %
style=authoryear,
natbib=true,        %
isbn=false,			%
url=false,          %
doi=false,
date=year,			%
giveninits=true,	%
sorting=nymt,		%
bibencoding=utf8,
maxcitenames=1,
]{biblatex}

\DeclareSortingScheme{nymt}{
	\sort{
		\field{presort}
	}
	\sort[final]{
		\field{sortkey}
	}
	\sort{
		\field{sortname}
		\field{author}
		\field{editor}
		\field{translator}
		\field{sorttitle}
		\field{title}
	}
	\sort{
		\field{sortyear}
		\field{year}
	}
	\sort{
		\field{month}
	}
	\sort{
		\field{sorttitle}
		\field{title}
	}
	\sort{
		\field[padside=left,padwidth=4,padchar=0]{volume}
		\literal{0000}
	}
}

\renewcommand*{\mkbibnamefamily}[1]{#1}  %

\DefineBibliographyExtras{french}{\restorecommand\mkbibnamefamily} %

\renewbibmacro{in:}{}   %

\DeclareFieldFormat
[article,inbook,incollection,inproceedings,patent,thesis,unpublished]
{title}{#1\isdot}       %

\DeclareFieldFormat[article]{number}{(#1)}    %

\AtEveryBibitem{%
	\clearlist{language}%
}

\AtEveryBibitem{%
	\ifentrytype{book}{%
		\clearfield{pagetotal}%
	}%
}

\usepackage{authblk}

\usepackage{tocloft}
\usepackage[english]{babel}
\usepackage[utf8]{inputenc}

\usepackage{graphicx,amsfonts}
\usepackage{amsmath}
\usepackage{verbatim}
\usepackage{textcomp}
\usepackage{amssymb}
\usepackage{pifont}
\newcommand{\xmark}{\ding{55}}
\usepackage{wrapfig}

\newtheorem{theorem}{Theorem}

\usepackage{empheq} 

\usepackage{siunitx}

\usepackage{tabu}
\usepackage{multirow}

\usepackage{enumitem}
\setlist{noitemsep}

\usepackage[title]{appendix}

\usepackage{color}

\definecolor{darkblue}{rgb}{0,0,0.5}
\definecolor{darkred}{rgb}{0.5,0,0}

\numberwithin{equation}{section}

\newcommand{\be}{\begin{equation}}
\newcommand{\ee}{\end{equation}}

\newcommand{\lp}{\left(}
\newcommand{\rp}{\right)}

\newcommand{\lbk}{\left[}
\newcommand{\rbk}{\right]}

\newcommand{\di}{\mathrm{d}}

\renewenvironment{quote}
{\small\list{}{\rightmargin=0cm \leftmargin=1.5cm}%
	\item\relax}
{\endlist}

\usepackage{hyperref}

\usepackage[nameinlink]{cleveref}

\begin{document}

\title{The concept of velocity in the history of Brownian motion}
\subtitle{From physics to mathematics and back}
\author[1,*]{Arthur Genthon}
\affil[1]{Gulliver, ESPCI Paris, PSL University, CNRS, 75005 Paris, France}
\affil[*]{arthur.genthon@hotmail.fr}
\date{\normalsize This is a post-peer-review, pre-copyedit version of an article published in The European Physical Journal H. The final authenticated version is available online at: https://doi.org/10.1140/epjh/e2020-10009-8}

\maketitle

\begin{abstract}
	{Interest in Brownian motion was shared by different communities: this phenomenon was first observed by the botanist Robert Brown in 1827, then theorised by physicists in the 1900s, and eventually modelled by mathematicians from the 1920s, while still evolving as a physical theory. Consequently, Brownian motion now refers to the natural phenomenon but also to the theories accounting for it. There is no published work telling its entire history from its discovery until today, but rather partial histories either from 1827 to Perrin's experiments in the late 1900s, from a physicist's point of view; or from the 1920s from a mathematician's point of view. In this article, we tackle the period straddling the two `half-histories' just mentioned, in order to highlight continuity, to investigate the domain-shift from physics to mathematics, and to survey the enhancements of later physical theories. We study the works of Einstein, Smoluchowski, Langevin, Wiener, Ornstein and Uhlenbeck from 1905 to 1934 as well as experimental results, using the concept of Brownian velocity as a leading thread. We show how Brownian motion became a research topic for the mathematician Wiener in the 1920s, why his model was an idealization of physical experiments, what Ornstein and Uhlenbeck added to Einstein's results, and how Wiener, Ornstein and Uhlenbeck developed in parallel contradictory theories concerning Brownian velocity.} 
\end{abstract}

\section{Introduction}
\label{sec:intro}
\noindent

Brownian motion is in the first place a natural phenomenon, observed by the Scottish botanist Robert Brown in 1827. It consists of the tiny but endless and random motion of small particles, contained in pollen grains, at the surface of a liquid. It naturally interested botanists until Brown and some physicists brought it into the field of physics. The physicists built the first quantitative theories to account for this motion, culminating with Albert Einstein, Marian von Smoluchowski and Paul Langevin in the 1900s. In the 1920s, Brownian motion knew a second domain shift, to mathematics, with Norbert Wiener's early works; while continuing to be studied by physicists like Leonard Salomon Ornstein and George Eugene Uhlenbeck. 

Although Brownian motion is well documented in the literature, its history is often split into parts that prevent us to appreciate its continuity and especially the transfers between disciplines. 
We can easily find excellent reviews of Brownian motion from a physical point of view\footnote{See  \citep{nye_molecular_1972,brush_kind_1976,maiocchi_case_1990} and \citep{duplantier_brownian_2007} which is a reviewed and extended version of the original paper \citep{duplantier_brownian_2006}}, starting in 1827 and usually ending around 1910, when physicists succeeded in building satisfactory theories, which were in addition confirmed by Jean Perrin's experiments.
After the 1910s, the `second-half' of the Brownian motion history started with ground-breaking progress made by Norbert Wiener from the 1920s and with the numerous enhancements to the existing physical theories made by Ornstein from the late 1910s onward, later joined by Uhlenbeck. 
This second history is hardly-ever told, or is written in difficult mathematical language\footnote{See \citep{kahane_essai_1998}.}. In any case, the continuity between the two half-histories is almost never dealt with. As a result of this assessment, the goals of this article are the following. 

We aim to fill the gap between the two half-histories of Brownian motion by asking how an object of interest for physicists could become a research topic for a young visionary mathematician; what were the points Ornstein wanted to work on in order to enhance the physical theory of Brownian motion when it already seemed successful at that time; and how do these two theories, developed in parallel, compare?

As a thread throughout this history, we chose to center the discussion on the concept of velocity, because of its central importance in both the physical and mathematical theories of Brownian motion, as well as in their comparison.
Indeed, the velocity of Brownian particles was one of the most difficult concepts to agree on for experimenters and theorists in the 1900s, and therefore was a debated topic that shaped the theory we know today. 
Secondly, the clarification of the notion of velocity at short-time scales was the starting point of Ornstein's later work and the main enhancement he brought to physical theories from the 1900s. 
Thirdly, the understanding of Perrin's account of irregular trajectories without a well-defined velocity was one of Wiener's primary motivations, and also a leitmotif of his entire work on Brownian motion, culminating with the non-differentiability of Brownian trajectories. Following this theme in his work will also be for us the occasion to explain some simple results of Wiener's theory to physicists, which are difficult to read in original form, though useful to understand the birth of the field of stochastic processes. 
Finally, the existence of velocity is a conflicting point between the physical and mathematical theories, which offers an illustration of how physicists and mathematicians can work on the same subject at the same time without truly communicating. This conflict led to a polysemy of the term Brownian notion, which refers today to both the natural phenomenon, and the physical and mathematical theories accounting for it. We hope that the detailed analysis of both theories will help the reader to disambiguate this term.

We start by giving a short review of Brownian motion history from 1827 to 1905, to set important landmarks and describe the context in which Einstein published his first article. 
Secondly, we sum up the history from 1905 to 1910, which includes the theories proposed by Einstein, Smoluchowski and Langevin, the experiments carried out by Theodor Svedberg, Max Seddig, Victor Henri and Perrin and the debates between the two communities. 
These elements clearly set, we have all the required background knowledge to study in detail in a third part Wiener's work from 1921 to 1933, in a fourth part Ornstein's and Uhlenbeck's works from 1917 to 1934 (with a glimpse at the 1945 article), and to finally compare these theories. 
We restrict ourselves to the study of texts published, translated or commented in English or French. 

\section{Historical background}
\label{sec_context}
\noindent
We aim to give in this section a quick review\footnote{For in-depth studies on this period, one can read \citep{brush_kind_1976,maiocchi_case_1990,duplantier_brownian_2007}.} of the period ranging from 1827 to 1905, during which sparse progress was made, to explain the context in which Einstein and Smoluchowski published their first articles and to offer the reader useful clues for the understanding of later works. Moreover, one key aspect of the works conducted during this period, and the debate that arose, is the interpretation of Brownian motion as a consequence of the atomic hypothesis, which brought the question of Brownian velocities to light. Indeed, the values of Brownian velocities were predicted by the atomic hypothesis and therefore served as a good testing quantity in the ongoing debate.\\

Brown was not the first to observe Brownian motion, but he was the first to repeat the experiment of observing a strange, irregular and endless motion for various suspended particles, including inorganic ones\footnote{In fact, Brown himself cited a 1819 work on this point, by Bywater from Liverpool, as related in \citep{duplantier_brownian_2007}, but denied the construction of his experiment.}. Doing this, he put an end to the vitalist theories relying on the hypothetical vital force animating living particles, and thus explaining the motion. From that moment on, he aimed at eliminating some physical explanations to this movement, such as the evaporation-induced fluid flows, or the interaction between suspended particles, with success.

The experiments conducted between Brown and Einstein were incomplete and too qualitative, thus leading to diverging interpretations. The authors neither agreed on the origin of the motion nor on the experimental results themselves. 

Concerning the origin of the motion, Christian Wiener, Louis Georges Gouy, P\`ere Julien Thirion, Ignace Carbonnelle and others invoked the kinetic theory of gas, introduced by James Clerk Maxwell and Ludwig Boltzmann, which we call the atomic hypothesis and which states that the velocity of Brownian particles is communicated by collisions with the medium particles. 
In 1863, Christian Wiener was the first one to propose a version of the atomic hypothesis \citep{nelson_dynamical_1967}, though a primitive version of it, formulated in terms of an ether and prior to Maxwell's version \citep{brush_kind_1976}. The atomic hypothesis encountered several difficulties at that time. In 1879, Karl Wilhelm von Nägeli, a botanist who had the advantage of being familiar with the kinetic theory of gases and knowing the orders of magnitude of masses and speeds, proposed a counter argument to the atomic hypothesis. He developed a theory of displacements of small particles of dust in the air and calculated that given the mass ratio between a gas molecule and a dust particle, the speed communicated by the collision between the two would be much too weak to explain the velocities observed experimentally. Some problems were also raised by defenders of the atomic hypothesis, like Gouy, who published in 1888 an article in which he recognized that uncoordinated collisions would not be enough to account for the motion of suspended particles, and thus a correlation would be needed on a space of about one micron. He was not the first one to make this observation, but being a physicist he was able to bring this difficulty into the physicists community, and hence was often wrongly presented as the discoverer of the origin of Brownian motion. Gouy's major contribution was to point out that the atomic hypothesis seemed to violate the second principle of thermodynamics, as the thermal energy from molecular agitation was converted into mechanical work providing velocity to suspended particles.
On the other hand, other physicists claimed that the motion was caused by various phenomena like lightning or electricity. Those last fanciful theories were refuted, at least qualitatively, during the twentieth century.

Concerning the experimental results, for most physicists the motion was truly random, whereas it was a deterministic oscillatory movement for Carbonnelle and Svedberg, as discussed in \cref{sec_exp}, for example. The influences of different parameters were also called into question. For Gouy and the Exner family, temperature increased the motion, in the sense that it increased the velocity of Brownian particles. Indeed, Siegmund Exner published in 1867 his observations indicating that the intensity of movement seemed to increase with the liquid's temperature and also when the liquid's viscosity decreased \citep{pohl_theory_2006}. The first quantitative and repeated measurements to study the influence of the particle's size and of the temperature on the velocity of suspended particles were carried by Felix Exner, Siegmund Exner's son, in 1900. He obtained an affine relation between the mean square velocity and the temperature, intersecting the axis at $T=\SI{-20}{\degreeCelsius}$, whereas the kinetic theory predicted a proportional relation. He himself had no opinion on his result, and did not see it as an argument in favour of a theory or another \citep{maiocchi_case_1990}.
On the other hand, for Thirion and Carbonnelle the opposite relation between velocity and temperature was true\footnote{On this point, Perrin later showed that the influence of the temperature had never been truly measured since viscosity also depends on temperature, therefore experimenters rather measured the influence of viscosity.}. 

Several factors were invoked to explain this lack of strong result during the nineteenth century, including the lack of interest of physicists for this phenomenon. Yet, the revival of the kinetic theory of gases and Maxwell's and Boltzmann's ground-breaking works in the years 1860-1890 boosted the researches on the links between microscopic theory and heat.
The lack of suitable mathematical tools was also highlighted for the first observations, since they occurred before or shortly after the 1860s, in which statistical methods from the kinetic theory of gases became available\footnote{Despite the link between the kinetic theory of gases and Brownian motion, it is striking to note that neither Maxwell nor Rudolf Clausius published on Brownian motion. Boltzmann was aware of some of the Brownian motion experiments, which he mentioned in a letter to Ernst Zermelo in 1896 \citep{darrigol_atoms_2018}, but he never tackled this issue, whereas it could have been a great test for his theory.}.
That said, according to Roberto Maiocchi, it is important not to take the lack of mathematical tools as solely responsible for the difficulties. As highlighted in this section, the set of experimental data was quite fuzzy so physicists were far from the ideal case where a strong set of cross-confirmed experimental data was only waiting to be accounted for by a theory.  
As we shall see, Einstein's theory on Brownian motion did not emerge from the knowledge of the experiments conducted in the nineteenth century, and more importantly, it concerned a quantity (displacement) that had never been measured during this period.

\section{Brownian motion as a physical concept}
\label{sec:theo_phys}
\noindent
After nearly 80 years without a satisfactory theory for Brownian motion, Einstein, Smoluchowski and Langevin published their works over a period of only four years, between 1905 and 1908. All three theories rely on the atomic hypothesis, which at that time was not yet accepted by the whole community. Those theories along with Perrin's experiments played a major role in its acceptance.

Einstein published a series of five articles on Brownian motion between 1905 and 1908, gathered in \citep{einstein_investigations_1926}. We decide to study the first one \Citep{einstein_uber_1905}, which contains all the ingredients of his theory and which is one of the reference article on the subject; and the fourth one in which he tackled the issue of the experimental measurement of Brownian velocities. The other three are not relevant for our study.

Smoluchowski started to work on Brownian movement before Einstein but he did not publish his results before 1906 \citep{smoluchowski_essai_1906}, as he was waiting for more experimental evidence and was finally pushed by Einstein's publication. He continued to work and publish on Brownian movement until his death in 1917.

Langevin wrote only one article on Brownian motion, in 1908 in the \textit{Comptes rendus de l'académie des sciences} \citep{langevin_sur_1908}.

Their articles have been much discussed in the literature\footnote{see \cite{brush_kind_1976,duplantier_brownian_2007,maiocchi_case_1990,nelson_dynamical_1967,nye_molecular_1972,piasecki_centenary_2007} for detailed analysis.} so we do not attempt to give a full rendition of their ideas but rather to highlight the main reasonings and the main results, because they are the starting point of Wiener's, Ornstein's and Uhlenbeck's works, as we shall see in the following sections. 

We propose to analyse each theory through three questions: (i) what are their physical ingredients? (ii) how do they introduce stochasticity into the equations? and (iii) what are their hypotheses concerning velocity?

We then look at the reception of these theories in the experimenters' world, through the works by Svedberg, Seddig, Henri and Perrin and the corresponding answers from Einstein, Smoluchowski and Langevin; because it offers some insights on the thorny understanding of Brownian velocity.

\subsection{The first quantitative theories}

\subsubsection{Albert Einstein - 1905}
\label{sec_einstein}
\noindent
In his 1905 article, Einstein obtained two major results: the relation between the diffusion coefficient and the properties of the medium; and the correspondence between Brownian motion and diffusion. Interestingly enough, it was not directly an article about Brownian motion since he declared that he did not know if the phenomenon he studied was what experimenters called Brownian motion, but it could be. His aim was not to account for experimental results but rather to propose a test for the validity of the kinetic theory of gases. If the kinetic theory of gases was true then microscopic bodies in suspension in a liquid should be in movement, and this motion should be observable with a microscope. On the other hand, such behaviour was forbidden by classic thermodynamics which predicted an equilibrium, thus putting the two theories in conflict. 
He then defined a measurable quantity which can weigh in favour of a theory or the other: the mean of the squares of displacements. 

Einstein's invention of the physical theory of Brownian motion was discussed in detail in \citep{renn_einsteins_2005}. In particular, J\"{u}rgen Renn analysed how Einstein combined his ideas coming from his 1901-1902 work on solution theory and his 1902-1904 work on the statistical interpretation of heat radiations, to come up with the idea that the atomic hypothesis could be tested by observing fluctuations from particles in solution. \\

Einstein's reasoning was built in three steps. In the first step, he related the diffusion coefficient to the properties of the medium, in a second step he derived the diffusion equation from a series of hypotheses on the particle's motion, and lastly he combined the two results.

Let us analyze the physical ingredients used in the first step, without going into details. Einstein used two physical ingredients from different validity domains, which was one of his master ideas. The first one is Stokes' law, which describes the force $\vec{F}$ that undergoes a spherical body of radius $a$ when in movement at constant velocity $\vec{v}$ in a fluid of viscosity $\mu$: $\vec{F}=-6\pi \mu a \vec{v}$. The second one is van 't Hoff law, similar to ideal gas law, which relates the pressure increase $\Pi$, called osmotic pressure and due to the addition of dilute particles in a solution; and the concentration $n$ of those dilute particles: $\Pi=nRT/N_A$, where $N_A$ is Avogadro constant, $T$ the temperature and $R$ the gas constant. Even if Einstein's theory was based on the atomic hypothesis and thus on the collisions between particles, it was not directly an ingredient he took into account in his calculations. 

First, Einstein considered the equilibrium between two force densities: the gradient of osmotic pressure, and an external force density (in this case the viscous force described to Stokes' law): $nF-\partial \Pi / \partial x=0$. 
Second, at equilibrium two processes act in opposite directions: a movement of the suspended particles under the influence of the force $F$, and a diffusion process considered as a result of the thermal agitation. This can be written by canceling the number of particles that cross a unit area per unit time due to both processes: $nF/6 \pi \mu a - D \partial  n / \partial x=0$.
By combining the two equilibrium relations with the definition of $F$ and $\Pi$, Einstein obtained the first\footnote{An Australian physicist named William Sutherland, derived a very similar equation in 1904, before Einstein. His equation was $D=\frac{RT}{N_A}\frac{1}{6 \pi \mu a}\frac{1+3\mu/\beta a}{1+2\mu/\beta a}$, where $\beta$ came from a generalized Stokes' law, and should be taken infinite to compare to Einstein's result. He presented his derivation in January 1904 in an Australian congress and published his result in the beginning of the year 1905 in the proceedings of the congress and then in March 1905 in \textit{Philosophical Magazine}, two months before Einstein's article. He was however completely forgotten for Einstein's benefit. To explain this historical curiosity, several hypotheses have been emitted like a misprint in his first article of 1905, Sutherland's weak influence in Europe or the chemistry-rooted style he used. For more details, see \citep{duplantier_brownian_2007,home_speculating_2005}.} and simplest form of the fluctuation-dissipation theorem, written as
\be
\label{eq_coef_dif}
D=\frac{RT}{N_A}\frac{1}{6 \pi \mu a} \,,
\ee
of which we can read the full derivation in \citep{duplantier_brownian_2007} (extended version of the original article \citep{duplantier_brownian_2006}). \\

In a second phase, he examined `the irregular movement of particles suspended in a liquid and the relation of this to diffusion'. To introduce the irregularity in his equations, he used probability distributions in the fashion of the kinetic theory of gases. The system Einstein studied is defined as follows. Particles are described by their positions $x$ in one dimension and undergo displacements $\Delta$ over a time $\tau$. 

The notion of displacement is central in Einstein's analysis, and is defined as the distance between two positions at different times. We note that he did not refer to the true length of the actual trajectory of a particle between two different times, thus the quantity $\Delta/\tau$ does not represent the true velocity of the particle. This subtle difference confirms that Einstein introduced a new quantity to describe Brownian motion, which had not been not discussed by experimenters in the nineteenth century ans which suited well the study of Brownian motion. This was a point of conflict with later experimenters as we will discuss in a moment. 

In his model, displacements are random and distributed according to a probability law $\phi_{\tau}$, normalised as $\int_{-\infty}^{+\infty} \phi_{\tau}(\Delta)\di\Delta =1$. Einstein called $f(x,t)$ the number of particles having a position between $x$ and $x+\di x$ at time $t$. $f$ is normalised at any moment $t$ as $\int_{-\infty}^{+\infty} f(x,t)\di x =N$, where $N$ is the total number of particles. Einstein next made a series of hypotheses:
\begin{enumerate}[label=(\roman*),align=left,leftmargin=1.75cm]
	\item Displacements of each particles are independent of that of others,
	\item We work at a timescale $\tau$ smaller than the observation time, but large enough for the displacements of a particle to be independent on two consecutive intervals of length $\tau$,
	\item The function $\phi_{\tau}$ is non-null only for small values of $\Delta$, in other words only small displacements are allowed over a time $\tau$,
	\item The space is isotropic, thus there is no privileged direction, and the probability distribution for displacements is even: $\phi_{\tau}(\Delta)=\phi_{\tau}(-\Delta)$.
\end{enumerate}
As we will see in a moment, future theories did not necessarily accept these hypotheses. Einstein nevertheless judged them natural and used them to write the relation between the distribution $f$ at time $t+\tau$ and that at time $t$ as follows
\be
f(x,t+\tau)=\int_{-\infty}^{+\infty}f(x+\Delta,t)\phi_{\tau}(\Delta)\di\Delta \,.
\ee
Using hypotheses (ii) and (iii), he expanded the left-hand side at first order in $\tau$ and the right-hand side at second order in $\Delta$, thus obtaining
\be
\label{eq:diffus_pos_phi}
\frac{\partial f}{\partial t} = \frac{\partial^2f}{\partial x^2} \cdot \frac{1}{\tau} \int_{-\infty}^{+\infty}\frac{\Delta^2}{2}\phi_{\tau}(\Delta) \di\Delta \,. 
\ee
Einstein recognised the diffusion equation\footnote{The diffusion equation had been established by Adolf Fick, in the continuity of Joseph Fourier work's on heat conduction and of Georg Ohm's work on electricity conduction.} \footnote{Einstein was not the first to establish the link between a random process and the diffusion equation. In fact Louis Bachelier, working under the direction of Henri Poincaré, published a memoir in 1900 \citep{bachelier_theorie_1900} in which he found the diffusion equation for options prices in market economy. His contribution to Brownian motion is studied in \Cite{dimand_case_1993}.} 
\be
\label{eq_diffus_pos}
\frac{\partial f}{\partial t} = D \frac{\partial^2f}{\partial x^2} \,,
\ee
for which he defined the diffusion coefficient as:
\be
\label{eq_ein_d_tau}
D=\frac{1}{\tau} \int_{-\infty}^{+\infty}\frac{\Delta^2}{2}\phi_{\tau}(\Delta)\di\Delta \,.
\ee
The solution to \cref{eq_diffus_pos} is
\be
\label{eq_distrib_pos}
f(x,t)=\frac{N}{\sqrt{4 \pi D t}}\exp{\lp -\frac{x^2}{4 D t} \rp} \,.
\ee
Einstein noticed that thanks to the independence described by his hypothesis (i), he could choose the starting point of each particle as the origin of the associated coordinate system, rather than a common one. Thus 
$f(x,t)$ becomes the number of particles having undergone a displacement $x$ between time $0$ and time $t$. The probability distribution for the displacements is naturally $f/N$.

Einstein computed the second moment of this distribution, which is the mean of the squares of displacements, as
\be
\label{eq_lambda_int}
\lambda_x^2=\langle x^2 \rangle=2Dt \,.
\ee
In the last phase, Einstein combined the results of the two first parts to obtain 
\be
\label{eq_ecart}
\lambda_x=\sqrt{\frac{RT}{ N_A}\frac{1}{3\pi \mu a}} \ \sqrt{t} \,.
\ee
This is probably the most famous result on Brownian motion, and Einstein presented it as a physically measurable quantity, that could be the test-quantity we talked about in the introduction of this section. From this perspective, he computed the numerical value $\lambda_x=\SI{0.8}{\micro\meter}$, taking $t=\SI{1}{\second}$, $T=\SI{17}{\degreeCelsius}$, $a=\SI{1}{\micro\meter}$ and $\mu=\SI{1.35e-2}{\pascal \second}$, which are typical values for Brownian motion experiments.\\

Before closing this section, let us take a few lines to mention a theoretical difficulty concerning the introduction of the timescale $\tau$, pointed out by \citep{ryskin_simple_1997}.
Timescale $\tau$ is defined between the microscopic timescale $\tau_{corr}$ for which there are correlations between displacements, and the macroscopic timescale $\tau_{macro}$ which is the characteristic time of variation for observable quantities, such as $f(x,t)$. Thus, $\tau$ cannot be taken to be $0$, however there are two steps in Einstein's calculation which implicitly suppose a $\tau \rightarrow 0$ limit: first, when doing the expansion in powers of $\tau$; second, in the identification of the diffusion coefficient with the integral term in \cref{eq_ein_d_tau}. Indeed, $D$ should not depend on an arbitrary timescale (besides, Einstein did note write $D_{\tau}$), whereas the right-hand side explicitly depends on $\tau$. The only escape from this contradiction is that in the $\tau \rightarrow 0$ limit, the right-hand side becomes independent of $\tau$.
How to satisfy both conditions? Gregory Ryskin argued that the limit is not formally reached, but people still write $\tau\rightarrow 0$ in the sense of $\tau \ll \tau_{macro}$. 

As we shall analyze, the supposition of the existence of $\tau$ and the lack of mathematical rigor in the treatment of its limit are weak points of Einstein's article, from which Wiener diverged from the physical Brownian motion (\cref{sec_maths}), and on which Ornstein and Uhlenbeck sharpened Einstein's theory (\cref{sec_theo_phys_2}).

\subsubsection{Marian von Smoluchowski - 1906}
\label{sec:smol}
\noindent
Smoluchowski published his article in 1906, pushed by the first two articles published by Einstein in 1905 and 1906. Einstein's and Smoluchowski's articles are very different in style. 

Firstly, Smoluchowski knew in details all experimental works carried on Brownian motion before 1906, he gave a clear account of them at the beginning of his article, and he constructed his theory in order to account for these observations; whereas Einstein was not sure that the problem he dealt with really was Brownian motion. 
Secondly, Smoluchowski's calculations were directly based on the collisions between particles, which was better than Einstein's approach in his opinion because it offered an intuitive understanding of the microscopic mechanism, even though both theories gave the same results. 
Thirdly, he introduced stochasticity by the mean of average quantities, while Einstein's computation was more general because it dealt with whole distributions containing more information than average quantities.
Lastly, he examined the case where the particle dimension was small compared to the mean free path of the solution's particles, whereas Einstein did not.
 
Smoluchowski must also be credited for the counter-argument by which he debunked Nägeli's criticism of the atomic hypothesis, evoked in \cref{sec_context}, and which had remained unanswered until 1906. His idea was that, even if the velocity communicated to a suspended particle by a collision is tiny (around $\SI{2e-6}{\milli\meter\per\second}$), as pointed out by Nägeli, one must not deduce that collisions are unable to move suspended particles at the measured velocities, if they act together. Indeed, even though the average position is null due to space isotropy, the mean of the deviation (a positive quantity)  from the initial position is non-null, and evolves as the square root of the number $n$ of collisions\footnote{See \Citep{duplantier_brownian_2007} for the complete derivation.}. 
Thus, if $n$ is large enough, most collisions cancel but $\sqrt{n}$ collisions contribute to a displacement in one direction. According to him, there are $10^{20}$ collisions per second in a liquid which makes $10^{10}$ collisions contributing to the displacement. Taking Nägeli's value for the velocity communicated by one collision, then $10^{10}$ collisions give the particle a velocity $\SI{e3}{\centi\meter\per\second}$. This value is false as well, because of voluntary simplifications made by Smoluchowski. Indeed, according to him, the absolute value of the change in velocity depends on the absolute value of the velocity before the collision, and is therefore different for each collision; and the probability of a collision that slows down the movement is greater than the probability of a collision that speeds it up. However, this was a victory against Nägeli's argument. 
This was only a qualitative answer for Smoluchowski who continued with a quantitative argument. The true value of the velocity is given by the equipartition of energy (\cref{eq_ere_smol}), and should therefore be $v=\SI{0.4}{\centi\meter\per\second}$, which is still not in agreement with the experimental values. In spite of this disagreement, this was the good value for Smoluchowski. 
Indeed, it is impossible to follow experimentally the true trajectory of a particle that undergoes $10^{20}$ collisions per second, therefore, observed trajectories are averaged trajectories for which the length of the path is greatly underestimated. The value $v=\SI{0.4}{\centi\meter\per\second}$ should therefore be the good one between two collisions, which is not a measurable timescale. \\

In Smoluchowski's model, the motion of Brownian particles was described by a random walk: suspended particle traveled on a straight line at constant velocity between two collisions, and when a collision with a particle of the medium occurred, the direction of the traveling particle was randomly re-defined. For the sake of simplicity, he considered that particles always traveled at their average velocity given by the principle of equipartition of energy, which may be written 
\be
\label{eq_ere_smol}
\langle v \rangle \sqrt{m}= \langle v'\rangle \sqrt{m'} \,,
\ee
where $v$ and $m$ are as usual the velocity and the mass of the Brownian particle and $v'$ and $m'$ are the velocity and the mass of the medium particles. Since Smoluchowski worked along way with average quantities, we stop writing brackets from now on. 
Moreover, he considered that Brownian particles were weakly deviated at each collision, by a constant small angle $\varepsilon=3v/4v'$.

If we note $\lambda$ the mean free path of medium particles, defined as the average distance medium particles travel in straight line before colliding with an other medium particle, and $a$ the Brownian particle radius, there are two cases: (i) $a<\lambda$ and (ii) $a>\lambda$. The second case is the most common one, experimented in the lab and described by Einstein, where suspended particles are significantly bigger than solution particles. \\

\setlength{\columnsep}{15pt} %
\setlength{\intextsep}{0pt}%
\begin{wrapfigure}[22]{r}{0.4\textwidth} %
	\begin{center}
		\includegraphics[width=\linewidth]{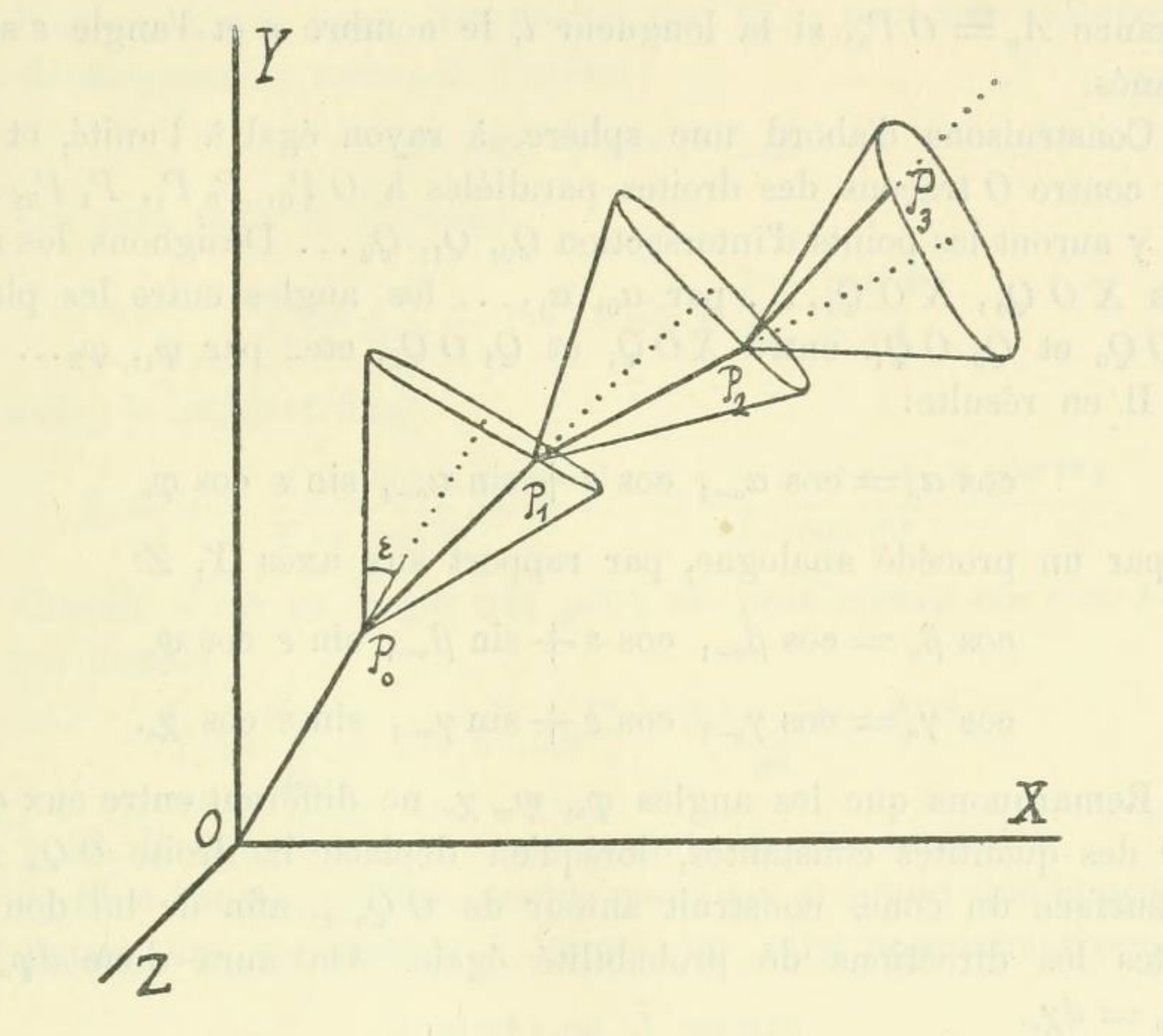}
	\end{center}
	\caption{Example of trajectory where $P_i$ are the collision points and $\varepsilon$ is the deviation angle. From \citep{smoluchowski_essai_1906}.}
	\label{fig_smol}
\end{wrapfigure}
Let us look at case (i) first. Smoluchowski made the asumption that the Brownian particle travels exactly the distance $l$ between each collision. Thus, the velocity of the particle, the distance it traveled between two collisions and the angle by which it is deviated when a collision occurs are constant; the only random parameter is the direction of the particle after a collision, for which Smoluchowski takes a uniform probability on the cone of angle $\varepsilon$. 

Like Einstein, Smoluchowski's understood the importance of the notion of displacement. Instead of aiming to compute the real distance traveled by the particle, which would be the sum of the lengths of all the segments between collisions, he defined the displacement undergone by the particle on a time $t$ as $\Lambda_n=OP_n$, where $O$ is the origin of the particle at time $0$ and $P_n$ is the position of the last collision at time $t$. This quantity is the straight distance between the to ends of the random walk. 
His goal was then to express the quantity $\sqrt{\langle \Lambda_n^2 \rangle}$, similar to Einstein's $\lambda_x$, as a function of the problem parameters. 
After some calculations, which can be found in \citep{duplantier_brownian_2007}, he demonstrated\footnote{In fact, there is a mistake in his calculations. Instead of the numerical factor $8/3$, he found $4\sqrt{2}/3$, but for the sake of clarity we choose to give the correct result.} that
\be
\label{eq_ecart_smol_partiel}
\sqrt{\langle \Lambda_n^2 \rangle}= \frac{8}{3} \, v'\sqrt{\frac{t}{n^*}} \,,
\ee
where $n^*$ is the number of collisions per second, defined by $n=t\cdot n^*$. The case (ii) is more difficult and gives a very similar result, so we choose not to analyse it, but rather to look at the comparison between the above result and Einstein's formula \cref{eq_ecart}. 

Smoluchowski sought a relation between the friction coefficient $S$ and the parameters $m$ and $n^*$, in order to compare his result to Einstein's one. According to him, usual methods gave $S=2m'n^*/3$, which allowed him to replace $n^*$ in \cref{eq_ecart_smol_partiel}. He also used Stokes' law $S=6\pi\mu a$ to substitute $S$ and finally obtained 
\be
\label{eq_ecart_smol}
\sqrt{\langle \Lambda_n^2 \rangle}= \frac{8}{9} \sqrt{\frac{m'v'^2}{\pi \mu a}} \, \sqrt{t} \,.
\ee
In order to compare this result to that of Einstein, we use the equipartition of energy to replace $m'v'^2$ by $k_B T$ for the one-dimensional case, although Smoluchowski did not do it, giving
\be
\sqrt{\langle x^2 \rangle}=\sqrt{\frac{64}{27}}\sqrt{\frac{k_B T}{3\pi \mu a}} \, \sqrt{t} \,,
\ee
which is Einstein's formula \cref{eq_ecart} with an additional factor\footnote{The additional factor was mentioned by Smoluchowski himself in his article, but with the value $\sqrt{32/27}$ due to the error of a factor $\sqrt{2}$ already discussed.} $\sqrt{64/27}$. This small difference is not surprising given all the approximations Smoluchowski made. 

Unlike Einstein, Smoluchowski wanted to compare his result with already existing data. He took for comparison Felix Exner's values, which he quoted in the introduction of his article, $ v = \SI{3.3e-4}{\centi \meter \per \second} $. According to \cref{eq_ecart_smol} with similar parameters to that of Exner, he got $ \sqrt {\langle x^2 \rangle} / t = \SI {1.3e-4} {\centi \meter \per \second} $. 
To compensate for this difference he introduced a rather mysterious coefficient $ \pi \sqrt{10} / 4 $ by which Exner's result must be divided. The $ \pi / 4 $ factor was a geometric correction to account for the fact that velocities were measured in a plane while the movement is three dimensional, but the $ \sqrt{10} $ factor is much more unjustified. By dividing Exner's value by this coefficient, he obtained $ v = \SI {1.33e-4} {\centi \meter \per \second} $, which was in agreement with his value.

At the end of his article, Smoluchowski obtained a relation between the diffusion coefficient $D$ and the parameters of the problem, by combining a qualitative reasoning on the mean free path and a result from his previous article on mean free path, which read 
\be
D=\frac{16}{243}\frac{m'v'^2}{\mu \pi a} \,.
\ee
This result is directly comparable to Einstein's in \cref{eq_coef_dif}, though still differing by a factor $64/27$.

\subsubsection{Paul Langevin - 1908}
\label{sec:langevin}
\noindent
In addition to his numerous personal contributions to physics, Paul Langevin is known to have read, understood and diffused Einstein's ideas on relativity and Brownian motion in France.
He published his only article on the subject in 1908 in the \textit{Comptes rendus hebdomadaires de l'académie des sciences} in full knowledge of Einstein's and Smoluchowski's articles. 

Langevin's derivation is so short and powerful, and radically different in the way he introduced randomness in the equations, that even if it has already been discussed in the literature, it is worth being given in details. 

Langevin started with the announcement of the exact correspondence between Einstein's and Smoluchowski's results, which differed until now by a factor $\sqrt{64/27}$, if one applies some corrections to Smoluchowski's derivation, even though he did not say which corrections.

Einstein (and also Smoluchowski in later articles we did not discuss) worked on probability distributions to establish partial differential equations which were deterministic, in the sense that they admit an exact solution, but for which the unknowns were the distributions, which are of probabilistic nature. On the contrary, Langevin used a probabilistic equation, which includes a stochastic noise and therefore cannot be solved directly, but which governs a deterministic variable: the velocity\footnote{Both ways of introducing the stochastic aspect of a problem in the equations are the pillars of the stochastic processes, as a branch of mathematics and theoretical physics. We still talk about Langevin equation (or stochastic equation) to refer to the case where a random variable appears in a partial differential equation, and about Fokker-Planck equation when the unknowns of the deterministic partial differential equation are probability distributions.
Fokker-Planck equation is of the same family as the one first used by Einstein and Smoluchowski, but was named after Adrian Fokker who worked with Max Planck on his thesis in 1913. His formula contains a convection term which makes it more general than the one applied to Brownian motion which only contains diffusion.} $\di x / \di t$. This equation, now known as Langevin equation, reads
\be
\label{eq_eq_langevin}
m\frac{\di^2 x}{\di t^2}=-6\pi \mu a \frac{\di x}{\di t} + X \,.
\ee
This is Newton's second law, applied to a particle subjected to a viscous frictional force governed by Stokes's law, and a stochastic force $ X$, whose origin is explained by Langevin as follows.
The viscous friction force only describes the average effect of the resistance of the medium, which is in reality fluctuating because of the irregularity of the collisions with the surrounding molecules. The stochastic force $X$ he introduced then accounts for the fluctuations around this average value. The two forces are therefore due to the same phenomenon: the collisions with medium particles, but one is averaged, deterministic and is in the opposite direction of the drift velocity, while the other is fluctuating, stochastic and has no privileged direction. Moreover, the value of $X$ is such that it maintains the particle's movement, which would stop otherwise because of the dissipative force.

Because of $X$, this equation cannot be solved exactly, so Langevin multiplied it by $x$ to obtain
\be
\frac{m}{2}\frac{\di^2x^2}{\di t^2}-m \lp \frac{\di x}{\di t} \rp ^2 = -3\pi \mu a \frac{\di x^2}{\di t}+xX \,.
\ee
He took the mean of the above equation over a large number of particles, making the term $\langle xX \rangle$ vanish because of the irregularity of the collisions\footnote{Langevin wrote `The mean value of $Xx$ is obviously null due to the irregularity of complementary actions $X$'. This physical intuition has been discussed and criticised in \citep{naqvi_origin_2005}.} described by $X$. He also replaced the term $m \langle (\di x/\di t)^2 \rangle$ by $RT/N_A$ using the equipartition of energy.  
Thus, he obtained a deterministic equation governing the newly-defined variable $z =\di \langle x^2 \rangle/\di t$, written as
\be
\frac{m}{2}\frac{\di z}{\di t}+3\pi \mu a z =\frac{RT}{N_A} \,.
\ee
This equation is similar to Einstein's type of equation discussed earlier since it is a deterministic equation governing a random variable, which is in this case not a probability distribution but the time derivative of one of its moments.

The solution is given by
\be
\label{eq_diff_lange}
z(t)=\frac{RT}{N_A}\frac{1}{3\pi \mu a}+C \exp{\lp -\frac{6\pi \mu a}{m}t \rp} \,,
\ee
where $C$ is a constant of integration. The second term in the right-hand side decreases exponentially with a characteristic time
\be
\label{eq_temps_carac_lang}
\theta=\frac{m}{6 \pi \mu a} \,,
\ee 
and thus becomes negligible when time $t>\theta$, whose value is approximately given by $\SI{e-8}{\second}$. This value is much smaller than measurable intervals so experimenters are always is the case where the first term in the right-hand side prevails. 
In this case, he replaced $z$ by its definition and integrated once to obtain the exact same result as Einstein's \cref{eq_ecart} for $\langle x^2 \rangle$.

In Langevin's analysis, the velocity is a key ingredient since it appears in Newton's second law by the mean of the viscous drag, whereas Einstein and Smoluchowski did not require velocity to exist because they instead used discrete stochastic models only relying on the displacement over a certain time. 
That said, Langevin's approach gives the same relation for $\langle x^2 \rangle$ in the case where $t$ is larger than $\theta$. In this case, the existence of velocity is just a step in Langevin's reasoning and disappears in the end: the suitable concept remains displacement. However, in the case where $t$ is smaller than $\theta$, the exponential term in \cref{eq_diff_lange} cannot be neglected and an extra exponentially-decreasing term appears in the relation for $\langle x^2 \rangle$. This has to be put alongside the existence of a timescale $\tau$ under which Einstein's derivation of the formula for $\langle x^2 \rangle$ does not hold.

\subsection{Experimental difficulties: From Svedberg to Perrin}
\label{sec_exp}
\noindent

Perrin's work has been deeply documented, and is often the logical following step in Brownian motion histories, after Langevin's theory. Therefore there is not much we can add to the existing literature, but we can instead describe preceding works by Svedberg, Seddig and Henri, since they are much less studied and because they are relevant to our understanding of the concept of Brownian velocity. 
Indeed, Svedberg's and Seddig's articles are characteristic of the misunderstandings on the notion of displacement introduced by Einstein and Smoluchowski. \\

Svedberg's articles have not been translated into English, thus the following biographical elements and analysis of his work come from \Citep{kerker_svedberg_1976} and \citep{sredniawa_collaboration_1992}; the discussion on the other physisicists' reactions can be found in \Citep{kerker_svedberg_1976} for Einstein and Perrin and in \citep{sredniawa_collaboration_1992} for Smoluchowski.

Theodor Svedberg was a Swedish physicist who studied Brownian motion thanks to Richard Adolf Zsigmondy's ultramicroscope invented in 1902, which he built himself with the help of Zsigmondy's plans. He published his results in 1906 \citep{svedberg_uber_1906},without any knowledge of Einstein's or Smoluchowski's works. In the same way as Einstein's article was not an attempt to account for previously existing data, Svedberg's measurements were not an attempt to test the theories. He then held two false ideas, first that he was able to measure the true velocities of suspended particles, and second that Brownian motion was oscillatory. The second mistake probably came from his lecture of Zsigmondy's work, who himself described Brownian motion as oscillatory, sometimes with an additional linear movement when the suspended particles were small enough. It appears that Svedberg thought that the oscillatory movement was the true Brownian motion and that the linear movement was an artefact that should be eliminated by setting proper experimental conditions. His goal was to determine the period and the magnitude of this oscillation and to deduce the true velocity of particles. He thus designed a quite ingenuous experiment in which particles were carried by a flowing liquid in a particular direction at constant velocity. He then described their movement around their equilibrium position through a sinusoid and collected his results in his 1906 article and tested the influence of the viscosity and of the particle size on the magnitude of the oscillations. According to his data, the velocity stayed quite stable at the value $v=\SI{0.03}{\centi\meter\per\second}$. even when varying the viscosity and the particle size. 
 
Svedberg later discovered Einstein's 1905 article but did not understand it and tried to connect his misconceptions on Brownian motion to Einstein's results in a second paper \citep{svedberg_uber_1906-1}. He replaced, in Einstein's \cref{eq_ecart}, $\sqrt{\langle \lambda_x^2 \rangle}$ by four times the amplitude of his supposed sinusoid, whereas the first quantity is stochastic and the second one is deterministic; and replaced time $t$ by the period of the oscillations, whereas the first one is non-specified and the second one is a property of the motion. These two confusions show the misunderstanding of Einstein's work in the experimental world in the first years. Svedberg then checked his data against the new formula and the results differed by a factor 6 or 7, which he judged tolerable. 

It is striking that in spite of this accumulation of mistakes, which Einstein, Langevin and Perrin soon remarked on, Svedberg continued to trust his theory all his life and was even awarded the chemistry Nobel prize in 1926, the same year Perrin received the physics Nobel prize, both for their contributions to Einstein's theory. Let us look at how physicists reacted to Svedberg's experiments.

Perrin was the most severe regarding Svedberg's theory and several signs of his criticism can be found in work. Here is a sample:
\begin{quote}
	Until 1908, there had not been published any verification or attempt that gave a clue about Einstein's and Smoluchowski's remarks. [Then in footnote] Svedberg's first work on Brownian motion is no exception [\cite{svedberg_uber_1906,svedberg_uber_1906-1}]. Indeed:
	\begin{enumerate}
	\item The lengths given as displacements are 6 to 7 times too high, which, supposing they are correctly defined, would be no progress, especially on the discussion due to Smoluchowski;
	\item Much more gravely, Svedberg thought that Brownian motion became \textit{oscillatory} for ultra-microscopic particles. It is the \textit{wavelength} (?) of this motion which he measured and used as Einstein's displacement. It is obviously impossible to test a theory taking as a starting point a phenomenon which, supposed exact, \textit{would be in contradiction with this theory}. I add that, at no scale Brownian motion shows an oscillatory behaviour.
	\end{enumerate}
	(\cite{perrin_les_1913}, p. 178-179)
\end{quote}

The most mysterious reaction was surely that of Smoluchowski, as documented in \citep{sredniawa_collaboration_1992}. When Smoluchowski's 1906 article arrived in Uppsala, where Svedberg worked, the latter had already published his second 1906 article, in which he compared his results to Einstein's formula. However, due to the numerical error in Smoluchowski's article, discussed in \cref{sec:smol}, Svedberg's result were closer to Smoluchowski's predictions than to that of Einstein. He thus wrote a letter to Smoluchowski to show him his results, to which the latter replied enthusiastically. From that moment on, the two physicists started a correspondence. Thanks to the help of Smoluchowski who suggested some small modifications, Svedberg published another article in 1907, in which his results were only 3 to 4 times too large compared to theoretical values, against 6 to 7 times in his 1906 article. Their scientific collaboration spanned the period from 1907 to 1914, including works on Brownian motion as well as on density fluctuations. Up until 1916, Smoluchowski cited Svedberg's articles on Brownian motion, even after Perrin's results published in 1908. 

In his only article, Langevin briefly criticised Svedberg's results for two reasons. First, his values differed by a factor $1/4$ from the theory, thus most likely speaking of his 1907 article. Second, he claimed that Svedberg did not measure the good quantity, which is $\langle x^2 \rangle$.

Einstein's answer is surely the most interesting for us because it offers some insights into the misuse of his concept of displacement for experimental purposes. It has been briefly discussed in \citep{kerker_svedberg_1976}, but we aim to give a more detailed analysis of the mathematical arguments Einstein gave to highlight why the experimental measurement of Brownian velocity is in fact not possible.  He published in 1907 his fourth article on Brownian motion \citep{einstein_theoretische_1907}, which opened with a reference to Svedberg's works, and in which he wished to clarify some theoretical points for experimentalists. He started from the equipartition of energy, written as $m \langle v^2 \rangle = 3RT/N_A$ and used Svedberg's values for temperature and particles mass to compute the square root of the mean squared velocity as 
\be
\label{eq_val_vit_ein}
\sqrt{\langle v^2 \rangle}=\SI{8.6}{\centi\meter\per\second} \,.
\ee
He then questioned the possibility to observe such a gigantic velocity. He used a simple reasoning  to show that it is in fact not possible. If one takes the simplified model in which the particle is only submitted to the frictional force, the equation governing the evolution of its velocity is then $m\di v/\di t=-6\pi \mu a v$ and the velocity decreases exponentially. 
Einstein computed the time $\theta$ after which the velocity is only $10\%$ of its initial value, as 
\be
\theta=\frac{m\ln(10)}{6 \pi \mu a} \,,
\ee
which for Svedberg's parameters takes the value 
\be
\theta = \SI{3.3e-7}{\second} \,.
\ee
This timescale is clearly not accessible experimentally, therefore it is not possible to observe the value of velocity given by \cref{eq_val_vit_ein}.
Moreover, Einstein considered a simplified case but in reality one has to take collisions into account, which makes the measure even more impossible. Indeed, for the mean velocity to be maintained at equilibrium according to the equipartition of energy, the velocity decrease due to viscosity must be balanced by collisions which transfer impulses to the particles. Since collisions are extremely frequent, the particle movement is altered even during the short timescale $\theta$, which makes it impossible to define a velocity\footnote{Einstein did not do it, but we can picture the number of collisions in question with the help of Smoluchowski's value given in \Citep{smoluchowski_essai_1906}. According to the latter, there are $10^{20}$ collisions per second in a liquid, so during the time $\theta$ for which the particle loses $90\%$ of its velocity, the particle undergoes $\num{3.3e13}$ collisions. It is therefore impossible to assign neither a value nor a direction to the velocity of the particle at this timescale.}.

At the end of his article, Einstein gave a more theoretical argument to explain that velocity is not a suitable quantity to describe Brownian motion. Using his \cref{eq_ecart}, he defined a quantity which would have the meaning of the average velocity of a particle during a time $t$, expressed as
\be
\frac{\lambda_x}{t} \propto \frac{1}{\sqrt{t}} \,,
\ee
This average velocity is proportional to the inverse of the square root of the experiment duration $t$, and thus does not reach any limiting value as $t$ decreases\footnote{Einstein already mentioned this idea at the end of his 1906 article on Brownian motion \citep{einstein_zur_1906}, in a section named `On the limits of application of the formula for $\sqrt{\langle \Delta^2 \rangle}$'. He defined the same quantity $\lambda_x/t \propto 1/\sqrt{t}$ diverging as $t \rightarrow 0$, which is physically impossible. Einstein explained it by the fact that an hypothesis he used when deriving his result is caught off guard when taking the limit $t \rightarrow 0$: the independence of collisions. Therefore, velocity values obtained by this calculation bear no meaning.}, while remaining larger than $\theta$. Thus, the value of this mean-velocity-like quantity has no meaning since it depends on the observation time. Therefore all velocities that are measured experimentally, since they are mean velocities by nature because of the experimental incapacity to follow the true path, are doomed to be dependent on the measurement time. \\

Seddig's case is more subtle since his misconceptions are less obvious. His work has not been translated into English neither, but was discussed in \citep{maiocchi_case_1990}, in which the following information can be found. He knew Einstein's works when he published in 1907 and 1908 his articles, in which he seemed to check the relation $\lambda_x\propto \sqrt{T/\mu}$, when varying $T$ with $t$ constant. Once again, Perrin later said that it was difficult to draw conclusions from these experiments, because the viscosity $\mu$ also depends on the temperature.
We must wait 1911 for Seddig to add details to his experiments from 1907 and 1908. From these new details it appears that he misunderstood $\lambda_x$ for the actual length traveled by particles during a time $t$ and not the displacement. To measure the length of the path, he tried to take long exposure pictures but this was too complex since the light required for the picture brought energy to the liquid and then distorted the results.  
For the sake of his experiment, he was therefore forced to send only two very close flash lights and to measure the distance traveled in straight line during these two flash, which was in fact the good reading of the quantity $\lambda_x$ although he was unaware of it. 
He tried to find a way to recover the true length of the path, which he thought to be the true meaning of $\lambda_x$, from the displacement, but never succeeded, leading to the publication of his results which are in agreement with theory. \\

Perrin's name is associated with the experimental verification of Einstein's results, but in fact he was interested in statistical physics questions, close to Brownian motion, even before reading Einstein's articles. 
In 1906, Perrin published an article unrelated to Brownian motion, in which he spoke for the first time of the interest for physics of mathematicians' functions without tangents \Citep{brush_kind_1976}. These functions are useful as an analogy for Perrin to describe the discontinuity of matter. Indeed, even if matter seems smooth and continuous it is in fact heterogeneous and discontinuous when looked through a microscope. 
This mathematical concept was later used again by Perrin to describe Brownian trajectories, which was a starting point of Wiener's work, as we shall se in \cref{sec_motiv_wiener}.
 
On May 11, 1908 Perrin published his first results on Brownian motion in the \textit{Comptes rendus de l'académie des sciences} \citep {perrin_agitation_1908}.
At first sight, this article was very surprising because Perrin announced that he verified Einstein's theory, but there was no sign of Einstein's work in this article. Perrin rather tested the altitude distribution of particles suspended in a liquid, for which he obtained an exponential distribution, and from which he stated the validity of Einstein's hypothesis. In 1909, Perrin admitted that when carrying the experiments he had no knowledge of Einstein's work, and what he called Einstein's hypothesis seems in fact to be the equipartition of energy. Since the equipartition of energy did not explicitly appear in Einstein's 1905 article, it is very likely that Perrin was only aware of Langevin's version of Brownian motion (published March 9, 1908), in which the equipartition of energy was highlighted.

On May 18, 1908, only one week after Perrin's article, Victor Henri, another French physicist working on Brownian motion at the same time but independently, published his account on the question \citep{henri_etude_1908}. Unlike Perrin, Henri knew Einstein's work, understood it and he proposed the first\footnote{Indeed, Seddig tested it in 1907, but as we saw he misunderstood the quantity $\lambda_x$, whereas Henri did not.} experimental test of \cref{eq_ecart}, which links the mean square of displacements to time and other parameters. He used a complex photographic set-up, working with two flash 0.05 seconds away, to test the formula. Unfortunately, he found that his results were 4 times larger than the ones predicted by the theory. This was a new failure for the atomic hypothesis, considering that this time the experiment and its interpretation were faultless.
On July 6, 1908 he published another article \Citep{henri_influences_1908}, which dealt a new blow to the theory. He found that the increase of the solution's acidity slowed down Brownian motion, whereas Brownian motion should only be impacted by one solution property: its viscosity, and viscosity was not changed by this small rise of acidity. 
No one ever detected errors or flaws in Henri's experiments, and thus no one could explain why his results were diverging from the theory. Perrin later obtained good results using the same method and declared
\begin{quote}
	The method was fully correct, and had the merit of being used for the first time. I do not know the cause that distorted the results. (\cite{perrin_les_1913}, p.180)
\end{quote}
On July 13, 1908 Jacques Duclaux took Svedberg's and Henri's experiments as an argument against the atomic hypothesis \Citep{duclaux_pressions_1908}. He particularly criticized the use of Stokes' law outside its domain of validity. Indeed, Stokes' law is supposed to be used for larger particles, around the millimetre, and the solution is supposed to be continuous, while neither of the two hypotheses is satisfied. Perrin answered this criticism on September 7, 1908 by publishing his conclusive test of Stokes' law validity at the scale of Brownian particles \Citep{perrin_loi_1908}. In reality, his reasoning was circular and was not a real proof, as demonstrated in detail in \Citep{maiocchi_case_1990}. However, the mistake was not revealed soon and Perrin's article scored a point.

Eventually, Joseph Ulysses Chaudesaigues, who was working in Perrin's lab on Brownian motion experiments at that time, published on November 30, 1908 the article that put a stop to the debate on the theory's validity \citep{chaudesaigues_mouvement_1908}. Perrin invented a protocol to prepare emulsions containing particles of the exact same size, which was a great advantage since it greatly reduced the uncertainty due to the particle size. Thanks to this particular method, they successfully tested \cref{eq_ecart} and also checked that the influence on the mean square of displacements of the particle size, the liquid viscosity, and the experiment duration were those predicted by Einstein's formula. Perrin's numerous experiments on Brownian motion from 1908 to 1913 are gathered in his 1913 best-selling \textit{Les Atomes}.

\section{Norbert Wiener's theory of Brownian motion}
\label{sec_maths}
\noindent

By the end of the 1910s, the physicists' Brownian movement reached a satisfactory stage of theorization since the theories proposed by Einstein, Smoluchowski and Langevin were experimentally confirmed by Perrin.
 Although some theoretical physicists continued to investigate the theory of Brownian motion, as we will see in the \cref{sec_theo_phys_2}, the next major results were obtained by mathematicians.

Norbert Wiener was a pioneer in the construction of the first rigorous and comprehensive mathematical model of Brownian motion from the 1920s. He remained ten years alone to be interested in a mathematization of Brownian motion but was then joined by many mathematicians, such as Raymond Paley and Antoni Zygmund with whom he collaborated from the 1930s, Andrei Kolmogorov, Joseph Leo Doob or Paul Lévy to name only a few.
At that moment, Brownian movement knew another domain shift, to mathematics, as it had passed from the hands of biologists to physicists in the nineteenth century.

In this section, we firstly raise the question of the reasons that led Wiener to propose a mathematical model for Brownian motion and how these reasons were related to the existence of the velocity of Brownian particles.
Secondly, we analyze how Wiener's primary motivations remained a thread in his construction, by surveying Wiener's work published between 1920 and 1933. Throughout this journey, we try to present key aspects of Wiener's theory in the most accessible way, while taking care of highlighting its continuity and its trajectory towards the study of the differentiability of the curves formed by Brownian trajectories.

\subsection{Norbert Wiener's motivations}
\label{sec_motiv_wiener}
\noindent
What were the reasons for Norbert Wiener, a young 25 years old mathematician, to publish his first article on Brownian motion in 1921, when it was not yet a subject for mathematicians? Perrin's description of Brownian trajectories by mathematicians' functions without tangent is often presented as the starting point of Wiener's interest in the question.

Indeed, Wiener spoke of Perrin's description of Brownian trajectories in his autobiography \textit{I am a mathematician - the later life of a prodigy} as follows
\begin{quote}
Here the literature was very scant, but it did include a telling comment by the French physicist Perrin in his book \textit{Les Atomes}, where he said in effect that the very irregular curves followed by the particles in the Brownian motion led one to think of the supposed continuous non-differentiable curves of the mathematicians. He called the motion continuous because the particles never jump over a gap and non-differentiable because at no time do they seem to have a well-defined direction of movement.	(\cite {wiener_i_1956}, p.38-39)
\end{quote}
Pesi Masani, Norbert Wiener's biographer, author of \textit{Norbert Wiener 1894 - 1964}, discussed Wiener's reading of Perrin's \textit{Les Atomes}, and referred in particular to the following quote
\begin{quote}
Those who hear of curves without tangents or of functions without derivatives often think at first that Nature presents no such complications nor even suggests them. The contrary. however, is true and the logic of the mathematicians has kept them nearer to reality than the practical representations employed by physicist (\cite{perrin_les_1913}, p.25-26, \cite {masani_norbert_1990}, p.79)
\end{quote}
which was `music to Wiener's ears'.
In his 1923 article, Wiener quoted Perrin, as translated by Frederick Soddy in 1910
\begin{quote}
One realizes from such examples how near the mathematicians are to the truth in refusing, by a logical instinct, to admit the pretended geometrical demonstrations, which are regarded as experimental evidence for the existence of a tangent at each point of a curve. (\cite{perrin_mouvement_1909}, p.81, \cite{wiener_differential_1923}, p.133)
\end{quote}
It is clear then that the mathematical hypothesis expressed by Perrin played a role in the birth of Wiener's interest in Brownian motion, but was this the only reason?
To answer this question, it is necessary to briefly study Wiener's biography, his connection, his mathematical interests and the publications preceding that of 1921. \\

The following biographical elements are taken from Wiener's biographies \citep{wiener_i_1956,masani_norbert_1990}.

Norbert Wiener was born in 1894 in the United States and entered Tufts University in Boston at only 12 to study mathematics and biology, obtained his A.B. degree in mathematics, and then entered Harvard Graduate School for Zoology in 1909. Unwilling to continue in this branch, he was transferred to Harvard Graduate School for Philosophy in 1911, where he studied philosophy and mathematics. In 1913, just 18 years old, he obtained his PhD in mathematical logic and went to study logic, philosophy and mathematics in Cambridge (England) with Bertrand Russell, thanks to a Harvard post-doctoral fellowship.

During his stay in Cambridge in 1913-1914, Russell advised him to open up to disciplines other than pure logic and mathematics foundations, and Russell mentioned the interface between mathematics and physics. Wiener followed his advice and read Rutherford's work on  electron theory, Niels Bohr's atomic theory, Einstein's and Smoluchowski's works on Brownian motion, and Perrin's \textit{Les Atomes}. It is interesting to note that neither Wiener nor Masani mentions reading Langevin's article, which may explain why all of Wiener's work was based on the Einstein-Smoluchowski approach; while Ornstein, Uhlenbeck and Doob took Langevin's formalism with stochastic noise as a starting point, as we shall see in \cref{sec_theo_phys_2}.

In Cambridge, Wiener was also attending mathematics lectures by Godfrey Harold Hardy, who would be the most influential teacher for the young Wiener. He discovered with Hardy other aspects of mathematics and especially Lebesgue integration, named after Henri-Léon Lebesgue. Russell's lessons also introduced him to Einstein's theory of relativity.
His interest in the interface between mathematics and the physical sciences arose at this time from his reading and from the influence of his two professors Russel and Hardy. However, Wiener did not decide to work on mathematical physics before 1921. What happened between 1914 and 1921 that led Wiener to study Brownian motion?

The period 1913-1919 was very scattered since he worked and studied successively in Göttingen, Columbia, MIT and Harvard and saw his activity disturbed by World War I. During these years, he focused on the foundations of mathematics and their structure, he then studied algebra, postulates systems and philosophy.

In 1919 he obtained a professorship at MIT, where he met Henry Bayard Philips, who `more than anyone else' introduced him to the physical aspect of mathematics with Willard Gibbs' work on statistical mechanics, which was a key element of his understanding of the role of statistics in physics.

Also in 1919, he inherited analytical mathematics books after the death of the mathematician Gabriel Marcus Green of Harvard, at that time his sister's husband. He began to read the fundamental works on analysis, which he had until now left aside, with a particular interest for Lebesgue's and Maurice Fréchet's works, the latter whom he later met at the 1920 Strasbourg congress.

His interest in probabilities came from another meeting, with Isaac Albert Barnett in 1919. Wiener said he asked him a mathematical subject to study and Barnett suggested to him the field of probabilities where random events were not points but curves. Indeed, at this time probability theory dealt only with discrete problems based on random variables, and there was no continuous probability theory, based on measure theory from mathematical analysis.
According to Wiener, `The world of curves has a richer texture than the world of points. It has been left for the twentieth century to penetrate into this full richness.' (\cite{wiener_i_1956}, p.36). 

Wiener thus spent a year trying to apply the Lebesgue integral to intervals whose points were themselves curves but it was too difficult. However, Wiener knew mathematician Percy John Daniell's work, who had formulated a new theory of integration in 1918. He then wrote an article in 1920 \citep{wiener_mean_1920} to propose developments on Daniell's theory in the direction of the integration on function spaces. This work was purely mathematical and had no direct link with Brownian movement but in 1920, Wiener read Geoffrey Taylor's work on turbulence and saw a perfect subject to apply the ideas developed in his 1920 article. Indeed, turbulence theory was based on average quantities depending on the whole movement. This attempt was a failure because the problem of turbulence was too tough to be solved so early, but Wiener knew another subject, distantly related to the problem of turbulence: Brownian motion.
\begin{quote}
Here I had a situation in which particles describe not only curves but statistical assemblages of curves. It was an ideal proving ground for my ideas concerning Lebesgue integral in a space of curves, and it had the abundantly physical texture of the work of Gibbs. It was to this field that I had decided to apply the work that I had already done along the lines of integration theory. (\cite{wiener_i_1956}, p.38)
\end{quote}
Thus, Wiener published an article in 1921 \citep{wiener_average_1921-1}, in which he applied the ideas of his 1920 article to Brownian movement, as he wanted to do for turbulence. From this moment, the study of Brownian trajectories, and the functions of these trajectories, became a guideline in Wiener's study of Brownian motion. This question was of different nature from those of the physicists, as explained in his biography
\begin{quote}
	The Brownian motion was nothing new as an object of study by physicists. There were fundamental papers by Einstein and Smoluchowski that covered it, but whereas these papers concerned what was happening to any given particle at a specific time, or the long-time statistics of many particles, they did not concern themselves with the mathematical properties of the curve followed by a single particle. (\cite{wiener_i_1956}, p.38) 
\end{quote}

\subsection{Norbert Wiener's pioneer work}
\label{sec_work_wiener}

Norbert Wiener is recognized as an immense twentieth century mathematician, for his contributions to the theorization of Brownian motion, the invention of Wiener's measure, his contributions to Gibbs' statistical mechanics and quantum physics, but especially his invention of cybernetics. In all his works, the style of mathematical reasoning developed during the study of Brownian motion from the 1920s is recognizable. 

We decide to divide his work published between 1920 and 1933 into three periods, for each of which he developed a different model of Brownian motion. 

The first period extended mainly between 1920 and 1922, during which Wiener developed his ideas on functional (i.e. functions depending on other functions and not points) averages, following Daniell's work. He developed an axiomatic theory of integration, not based on measure theory. We have already mentioned two articles written during this period in the section on Wiener's motivations  \citep{wiener_mean_1920, wiener_average_1921-1}, but there were two other articles, published in 1921 \citep{wiener_average_1921} and in 1922 \citep{wiener_average_1922}. The first one was extremely little quoted in the secondary literature (with the exception of \citep{doob_wieners_1966}, which did not however analyze the article in detail). This article, however, was in the logical continuity of the 1920 and 1921 articles but, unlike the two previous ones, was much closer to physical questionings and rich in lessons on Wiener's Brownian movement.
The 1922 article was the culmination of Wiener's ideas on axiomatic integration where he developed a model that would later be taken up and improved in his 1930 article \citep{wiener_generalized_1930}, which we analyse in the third sub-section.

The second period began in 1923 with the publication of \textit{Differential Space} \citep{wiener_differential_1923}, the article often cited as the foundation of Wiener's theory of Brownian motion. He developed a different approach from that used until now, based on measure theory, and defined his well-known Wiener measure. It was also in this article that Wiener gave the first argument for the non-differentiability of Brownian trajectories by defining a coefficient of non-differentiability.

Lastly, Wiener returned to the question of Brownian motion in 1930, in his memoir \citep{wiener_generalized_1930} on harmonic analysis. Based on the approach developed in the 1922 article \citep{wiener_average_1922}, he constructed a third model of Brownian motion, based on the Lebesgue measure. The mapping he invented in this article to relate the set of continuous functions and the interval $[0,1]$, thus allowing the use of Lebesgue measure, was later used in his 1933 article \citep{paley_notes_1933}, resulting from the fruitful collaboration with Paley and Zygmund during the years 1932 and 1933, where they gave the final proof of the non-differentiability of mathematical Brownian trajectories.

It is through these three periods that we perceive the mathematical edifice built by Wiener, from Perrin's famous hypothesis up to the non-differentiability of Brownian trajectories. This highlights the importance of the in-existence of Brownian velocities, which marked both the beginning and the end of Wiener's construction.

\subsubsection{Mathematical context}
\label{sec_wien_context}
\noindent
To fully understand what was at stake in Wiener's articles, it is important to make a naive and very quick point on the state of the art of integration in 1920. 

In the second half of the nineteenth century, the first rigorous theory of integration had been developed by Bernhard Riemann. This theory was fundamental but had limits, which we do not expose here but which pushed mathematicians to seek another approach to integration. Thus, in 1902, Lebesgue proposed his version of integration, which made it possible to integrate functions on more complex spaces than the intervals of $ \mathbb {R}^N $, as for example sets of discrete points. We write his integral $ \int f \di \mu $ where $ \mu $ is the Lebesgue measure that gives a weight to each subset of the integration space. There is no general expression for this measure, but it is the simplest one in the sense that on simple spaces it corresponds to the intuitive notion of measure. For example, the Lebesgue measure of a segment is its length, the Lebesgue measure of a surface is its area, and so on. These two versions do not directly allow integration on sets of functions. Moreover, both are measure-based theories, that is, based on the measure $ \di x $ or $ \di \mu $ which weighs each element of integration.

In order to generalize the notion of integration to infinite-dimensional spaces, Percy John Daniell proposed in 1918 an axiomatic theory of integration, not based on measure theory. He defined an abstract object $ I $, which satisfied some axioms so that $ I (f) $ represented the integral of the function $ f $, and coincided with the prior definitions of the integral under certain conditions.

Like the previous constructions (Riemann, Lebesgue), Daniell first defined his integral on a very small set $ T_0 $ of functions, then showed how to extend the definition to a much larger set $ T_1 $, in the same way Riemann first defined his integral on step-functions before defining it on the set of piecewise continuous functions using his step-functions.

In the series of articles we examine in the following section, Wiener took up the ideas of Daniell's integration theory, to explicitly compute functional averages over function spaces. Let us then expose the premises of Daniell's theory, which is important to study Wiener's way of thinking and to understand his major results. \\

Daniell defined two abstract objects $ (I, T_0) $. $ T_0 $ is a space of simple functions on which the integration is simply defined, and $ I $ is the integration operator on $ T_0 $. Thus Daniell's integral is noted in all generality $ I (f), \ f \in T_0 $. Functions in $ T_0 $  are required to have the following stability properties
\be
\label{eq_cond_t0_daniell}
\left\{
\renewcommand{\arraystretch}{1.5}
\begin{array}{l}
	
	\forall (f_1,f_2) \in T_0^2, \ f_1+f_2 \in T_0 \,, \\
	\forall f \in T_0, \ \forall c \in \mathbb{R}, \ cf \in T_0 \,, \\
	\forall f \in T_0, \ \left| f\right| \in T_0 \,.  \\
	
\end{array}
\right.
\ee
Similarly, to match the intuitive idea of integration, the $ I $ operator must satisfy the following axioms
\be
\label{eq_cond_i_daniell}
	\left\{
	\renewcommand{\arraystretch}{1.5}
	\begin{array}{l}
		
	\forall (f_1,f_2) \in T_0^2, \ I(f_1+f_2)=I(f_1)+I(f_2) \,, \\
	\forall f \in T_0, \ \forall c \in \mathbb{R}, \ I(cf)=cI(f) \,, \\
	f \geq 0 \ \Rightarrow \ I(f) \geq 0 \,, \\
	f_1 \geq f_2 \geq ... \geq f_n \rightarrow 0 \ \Rightarrow \ \lim\limits_{n \to \infty} I(f_n)=0 \,.
	
	\end{array}
	\right.
\ee

Daniell's major theorem is to extend the integrability of the functions of $ T_0 $ to a much larger class of functions $ T_1 $, defined by the functions of $ T_0 $.

\begin{theorem}
	Let there be an increasing sequence of functions $ f_n $ belonging to $ T_0 $, such that there exists a function $ g $ greater than all the functions $ f_n $, then the limit $ f $ of the sequence $ f_n $ is summable in the sense of Daniell, with
	
	\be
	I(f)=\lim_{n \to \infty} I(f_n) \,.
	\ee
\end{theorem}
The set $ T_1 $ is then defined as the set of functions $ f $ described in the theorem.

\subsubsection{Axiomatic theory of integration on functions sets - 1920-1922}
\label{sec_wiener_1920}
\noindent
The 1920 article, as discussed in \cref{sec_motiv_wiener}, was unrelated to Brownian movement but exposed Wiener's progresses on Daniell's integration, which were later used in the following articles \citep{wiener_average_1921-1,wiener_average_1922}, both dealing with Brownian motion. Therefore, we found useful to expose the framework laid in the 1920 article in a first time.

Wiener noted that Daniell had established a method to go from $ T_0 $ to $ T_1 $ but left open how to build $ I $ and $ T_0 $ in the first place. Wiener proposed in this article to build these two objects and to apply them to the case of functionals. For this, he used a simple notion of step functions for $ T_0 $. We present here his construction step by step.

Let $ K $ be a set, we call $ I_n $ a division of the set $ K $ depending on a parameter $ n $, a division of $ K $ being defined as a finite set of subsets, also called intervals, which cover $ K $ at least once. The intervals of the division $ I_n $ are denoted $ i_1 (I_n) $, ..., $ i_m (I_n) $. We no longer use the $ I $ notation for the Daniell integral, so there is no confusion with the divisions.
We can also assign a weight $ w_{I_n} $ (denoted $ w_n $ if there is no ambiguity) to each interval of a division $ I_n $, so that the interval $ i_j (I_n) $ has the weight $ w_n \lbk i_j (I_n) \rbk $. The division $ I_n $ is then said to be weighted by $ w $.
Finally, a sequence $ \left \lbrace I_n \right \rbrace_{n \in \mathbb {N}} $ of divisions weighted by $ w $ is called a partition $ P_K $ of the set $ K $ if it satisfies the following properties
\begin{enumerate}[label=(\roman*),align=left,leftmargin=1.75cm]
	\item \label{it_div_inclu} Each interval of $ I_{n + 1} $ is included in an interval of $ I_{n} $ and only one, 
	\item  The weight $w_n \lbk i_j(I_n) \rbk$ of an interval $i_j(I_n)$ is the sum of the weights $w_{n+1} \lbk i_l(I_{n+1}) \rbk$ of the intervals $i_l(I_{n+1})$ included in $i_j(I_n)$.
\end{enumerate}
There is a third condition that does not contribute anything to understanding, which we do not give here for the sake of synthesis.

With these definitions, Wiener could then define his step functions. A function $ f $ defined on $ K $ is called a step function on $ P_K $ if there is a division $ I_n $ belonging to the partition $ P_K $ such that $ f $ is constant on each interval $ i_j (I_n) $.
Then Wiener defined the average $ A_{P_K} (f) $ of a step function $ f $ on $ P_K $ intuitively as
\be
\label{eq_moy_esc}
A_{P_K}(f)= \frac{\sum_{j=1}^{m} w_n \lbk i_j(I_n) \rbk f(x_j)}{\sum_{j=1}^{m} w_n \lbk i_j(I_n) \rbk} \,,
\ee
where $x_j \in i_j(I_n)$.\\

Wiener then showed that his step functions satisfied the conditions of the set $ T_0 $, given in \cref{eq_cond_t0_daniell} by Daniell; and that his definition of the mean (\cref{eq_moy_esc}) satisfied the axioms of the $ I $ operator, given in \cref{eq_cond_i_daniell}.
Using Daniell's theorem, Wiener proved that all bounded and uniformly continuous functions on $ P_K $ are summable in the sense of \cref{eq_moy_esc}. He thus had a construction of the mean of a function defined on any set $ K $, potentially of infinite dimension.

To conclude his article, Wiener took some examples. By defining the $ I_n $ divisions in a simple way and taking a segment for $ K $, his definition of the average gave back Lebesgue's one.
More interestingly, when he took for $ K $ the set of continuous functions defined on the interval $ [0,1] $ and null in $ 0 $, which we note $ C_0 [0 , 1] $ from now on, which are in addition bounded and Lipschitzian (then the functions defined on $ K $ were functionals by definition), then the application of the theorem gave that all continuous and bounded functionals were summable in the sense of Wiener. \\

In the following articles, Wiener gave more explicit definitions for the mean of a functional and applied his axiomatic theory to the study of Brownian motion. The next article \citep{wiener_average_1921-1} was the first to explicitly deal with Brownian motion and was fundamental in the construction of Wiener's idealized Brownian motion, as we shall analyze now. 

Wiener acknowledged René Gateaux's work on the theory of functional average but claimed his own version was more adapted to the case of Brownian movement than that of Gateaux. He began his article with a reference to Einstein's work, and recalled this result: if a particle is free to move on the $ x $ axis and is subjected to Brownian motion, and if we assume that the probability that it moves a certain value over a certain time interval is independent of
\begin{enumerate}[label=(\roman*),align=left,leftmargin=1.75cm]
	\item its starting point,
	\item its starting absolute time,
	\item its direction,
\end{enumerate}
then Einstein showed that the probability that after a time $ t $ the particle reached the position $ x $, written $ f (t) $ by Wiener,  between $ x_1 $ and $ x_2 $, was under certain assumptions
\be
\label{eq_wien_prob}
P(x_1 \leq f(t) \leq x_2) = \frac{1}{\sqrt{\pi t}} \int_{x_1}^{x_2} \exp{\lp-\frac{x^2}{t}\rp} \di x \,,
\ee
where Wiener voluntarily omitted to note the physical parameter $ 4D $, present in \cref{eq_distrib_pos}, by setting it equal to 1. The assumptions in question were not explained by Wiener but it is clear that the one of the existence of a time scale $ \tau $ on which the displacements are independent of previous displacements, was essential to the establishment of the Gaussian probability for Einstein. Wiener did not deal with the question of this time scale and used \cref{eq_wien_prob} for all times, which made his object a simplified model of Brownian motion. In fact, he freed himself from the physical difficulties that appeared when the mean free path was approached, and constructed a mathematical model that made it possible to study Brownian motion by extending the range of validity of the Gaussian distribution. It is this model that he continued to use in his subsequent articles and which he described lucidly as follows
\begin{quote}
In the physical Brownian motion, it is of course true that the particle is not subject to an absolutely perpetual influence resulting from the collision of the molecules but that there are short intervals of time between one collision and the next. These, however, are far too short to be observed by any ordinary methods. It therefore becomes natural to idealize the Brownian motion as if the molecules were infinitesimal in size and the collisions continuously described. It was this idealized Brownian motion that I studied, and which I found to be an excellent surrogate for the cruder properties of the true Brownian motion. (\cite{wiener_i_1956}, p.39)
\end{quote}

Let us go back to the article of 1921. To meet the conditions of his previous article, Wiener restricted the parameter $ t \in [0,1] $. Thus the functions $ f (t) $, describing Brownian trajectories fell within the framework of the second example given in the previous article, and he could compute the average of the continuous and bounded functions defined on this set $ K $ of functions $ f (t) $. Wiener gave in this article a more explicit formula for the average of these functionals. \\
Let us recall that functionals are defined as functions of functions, i.e. functions which do not depend on a finite number of variables but of an entire function, which can be seen as an infinity of variables (i.e. $ f \equiv \{f (t) \}_{t \in [0,1]} $). Physically, these functionals of trajectories can be any quantity that depends on the complete trajectory, such as the maximum value of the function, which represents the maximum distance the particle has moved away from its origin; or the length of the trajectory.

Wiener started with a simple case where the functional $ F [f] $, which we note with brackets to differentiate it from a simple function, depended on $ f $ only for a finite number of values $ f (t_1 ) $, ..., $ f (t_n) $ in polynomial form: $ F [f] = f (t_1)^{m_1} ... f (t_n)^{m_n} $. In this case $ F $ was rigorously a function and not a functional, and the average of $ F $ was conventionally defined as the average of a function
\be
\label{eq_moy_func_pol}
A(F)=\frac{1}{\sqrt{\pi^n t_1(t_2 \! - \! t_1)...(t_n \! - \! t_{n-1})}} \int\limits_{-\infty}^{+\infty} \di x_1 ... \int\limits_{-\infty}^{+\infty} \di x_n \, x_1^{m_1} ... x_n^{m_n} \, \exp{\lbk- \sum\limits_{k=1}^{n} \frac{(x_k \! - \! x_{k-1})^2}{t_k \! - \! t_{k-1}}\rbk} \,.
\ee
These integrals are analytically computable. Let us turn now to the most general case of true functionals. Wiener defined a general functional $ F [f] $ by the expression
\be
\label{eq_functional def}
F[f]=a_0 + \int_{0}^{1} f(t)\di \psi_1(t) + ... + \int_{0}^{1} ... \int_{0}^{1}  f(t_1) ... f(t_n) \di \psi_n(t_1, ..., t_n) + ... \,.
\ee
It was natural to require from the average operation to be stable by permutation with the sum of a series, by permutation with an integral, and by multiplication by a constant, which led Wiener to define the mean on this functional class as
\be
\label{eq_moy_functional def}
A\{F\}=a_0 + \int_{0}^{1} A(f(t))\di \psi_1(t) + ... + \int_{0}^{1} ... \int_{0}^{1}  A(f(t_1) ... f(t_n)) \di \psi_n(t_1, ..., t_n) + ... \,,
\ee
where we take care to note $ A \{\cdot \} $ the average of a functional, defined by this formula, and $ A (\cdot) $ the classical average of a function.
The right-hand side is computable using  \cref{eq_moy_func_pol} for the terms $ A (f (t_1) ... f (t_n)) $.
As soon as this series converges, we have a definition for the mean of a functional and a method relying on means of functions to compute it. \\

From here one can read Wiener's work following two paths: the article \citep{wiener_average_1922}\footnote{The date of this article is uncertain, because its heading specified `Received February 27th, 1922.—Read March 9th, 1922.' and Wiener cited it in his 1930 article \citep{wiener_generalized_1930} with the date 1922. However, this article is mentioned with the date 1924 in the literature, as for example in \cite{masani_norbert_1990}. Furthermore, in this precise article Wiener cited his 1923 article \textit{Differential Space}, whereas in \textit{Differential Space} he mentioned a `forthcoming paper in Proc. Lond. Math. Soc.' which is the 1922/24 article in question. It is then likely that both articles were written almost at the same time, during the year 1922, but were published in two different journals one year apart.} followed the logic of the two articles that we just presented to refine the theory of the calculation of functional average, and \citep{wiener_average_1921} deviated from this problem to deal on a concrete case coming from Brownian movement. We choose to break the chronological order here to present the article \citep{wiener_average_1922} first, because of its thematic filiation with the previous articles. The article \citep{wiener_average_1921} deserves a sub-section (\cref{sec_wien_1921}) for itself because it was the only article closely related to the physicists' Brownian movement and to experiments, and moreover it has hardly been discussed in the literature.\\

The purpose of the 1922 article was to specify the partition $ P_K $ used in the case of Brownian motion, which had not been done in the article \citep{wiener_average_1921-1}, and to use this partition to express the average of a functional having for argument a function representing the trajectory of a Brownian particle.

Wiener chose once again the set $ K = C_0 [0,1] $ and remarked that the distribution given by \cref{eq_wien_prob} made it possible to assign weights to certain groups of trajectories which all passed through the same intervals at certain given times. Indeed, he set $ n $ time values between $ 0 $ and $ 1 $: $ 0 \leq t_1 \leq ... \leq t_n \leq 1 $ and chose values $ x_{i1} $ and $ x_{i2} $ defining $ [x_{i1}, x_{i2}] $ windows through which the function $ f $ had to pass at times $ t_i $. Constraints on $ f $ were then expressed as
\be
\label{eq_fenetres_x}
	\left\{
	\renewcommand{\arraystretch}{1.5}
	\begin{array}{r c l}	
			
	x_{11} & \leq f(t_1) \leq& x_{12} \,, \\
	x_{21} & \leq f(t_2) \leq& x_{22} \,, \\
	&...& \,, \\
	x_{n1} & \leq f(t_n) \leq& x_{n2} \,.
		
	\end{array}
	\right.
\ee
The probability of observing a trajectory satisfying these constraints therefore was
\be
\label{eq_wien_proba_intervalles}
\frac{1}{\sqrt{\pi^n t_1(t_2 \! - \! t_1)...(t_n \! - \!  t_{n-1})}} \int\limits_{x_{11}}^{x_{12}} \di\xi_1 ... \int\limits_{x_{n1}}^{x_{n2}} \di\xi_n \, \exp{\lbk -\sum\limits_{k=1}^{n} \frac{(\xi_k-\xi_{k-1})^2}{t_k-t_{k-1}}\rbk} \,.
\ee
Wiener called the set of trajectories that satisfied the constraints of \cref{eq_fenetres_x} an interval, and defined the weight of this interval as the probability given by \cref{eq_wien_proba_intervalles}. We directly see the link with the construction of the article \citep{wiener_mean_1920}, though Wiener let the reader make the link with the concepts defined in the latter, which are the following. The interval defined by \cref{eq_fenetres_x} is what was previously noted $ i $, the probability of this interval is the weight $ w (i) $, and $ n $ which represents the number of time points, that is to say the fineness of the $ [0,1] $ axis division, is the parameter $ n $ on which the $ I_n $ divisions previously defined depend. 

To pursue the calculations, Wiener set explicit values for $ t_i $, $ x_{i1} $ and $ x_{i2} $, as follows
\be
\label{eq_def_part_K}
	\left\{
	\renewcommand{\arraystretch}{1.5}
	\begin{array}{r c l}	
		
	1 \leq&h&\leq 2^n \,, \\
	t_h&=&h / 2^n \,, \\
	x_{h1}&=&\tan(k_h \pi / 2^n) \,, \\
	x_{h2}&=&\tan((k_h+1) \pi / 2^n) \,,
		
	\end{array}
	\right.
\ee
where $k_h$ was an integer between $-2^{n-1}$ and $2^{n-1} -1$. \\

\begin{figure}[!h]
	\begin{center}
		\includegraphics[width=0.9\textwidth]{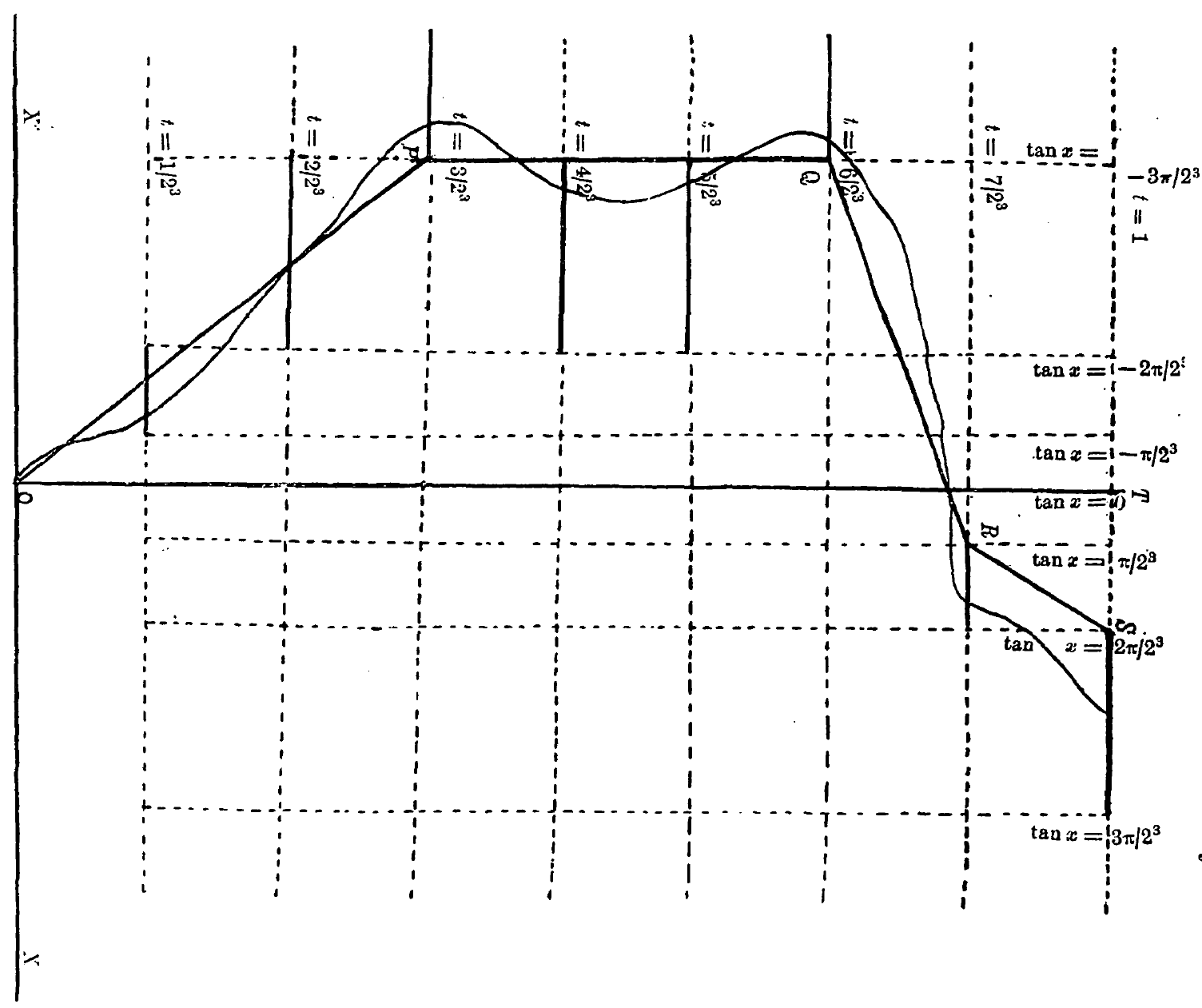}
		\caption{Example of a path going through the windows of a particular interval $ i (I_3) $. The image has been rotated by 90 ° compared to the one in the original article.}
		\label{fig_wiener}
	\end{center}
\end{figure}

Let us analyse these definitions. For a value of $ n $, we regularly split the time interval $ [0,1] $ with  $ 2^n $ time values $ t_h $. The first time value is $ t_1 = 1/2^n $ because the value of the function is set to $ 0 $ for $ t = 0 $. There are $ 2^n $ possible values for each $ k_h $, so $ k_h \pi / 2^n $ is ranging from $ - \pi / 2 $ to $ \pi / 2 $ and therefore $ x_{h1} $ and $ x_{h2} $ scan all $] - \infty, + \infty [$. A specific choice of the value of $ k_h $ for each $ h $ corresponds to an interval $ i (I_n) $. Since there are $ 2^n $ values of $ h $ and $ 2^n $ choices for $ k_h $, there are therefore $ (2^n)^{2^n} $ intervals in $ I_n $. An example of trajectory is presented on \cref{fig_wiener} for the division $ I_3 $. The curved line represents the Brownian trajectory, that is, a continuous function on $ [0,1] $ starting at $0$ for $t= 0 $. The horizontal dotted lines represent the possible values of the $ x_h $ and the vertical dotted lines represent the 8 possible values of $ t_h $. The black vertical segments highlight the windows through which the function must pass at each time $ t_h $. The choice of these windows is therefore an interval $ i(I_3) $ among the $ 8^8 $ possible.

It was relatively easy for Wiener to verify that this definition of the divisions $ I_n $, as well as that of the weight function $ w $ defined by the \cref{eq_wien_proba_intervalles} satisfied the properties (i) and (ii) established in his 1920 article \citep{wiener_mean_1920}. We just showed how Wiener established an explicit partition $ P_K $ on $ K = C_0 [0,1] $, which would be used in the later article \citep{wiener_generalized_1930} to redefine Brownian motion thanks to the Lebesgue integral. With the help of this partition, he could apply the results of the article \citep{wiener_mean_1920} to his set of functions.

Wiener followed the same construction: when in his 1920 article he built the step functions in order to define the mean and then used Daniell's theorem to extend this mean to uniformly continuous and bounded functions understood as limits of step functions series; in this article he defined the step functional in the same way and then extended the average operation to functionals that are limits of step functionals series. He defined a step functional as taking constant values on each interval $ i (I_n) $ of a certain division $ I_n $.
The average of a step functional was then naturally given by
\be
\label{eq_moy_fonctionnelle_esc}
A\{F\}=\sum_{k}F[f]w[i_k(I_n)] \,,
\ee
where $f \in i_k(I_n)$.

Wiener was finally able to apply Daniell's theorem and he deduced that all bounded and continuous functionals, defined on Brownian trajectories, were summable. Moreover the average value was given by the formula
\begin{multline}
\label{eq_moy_func_lim}
A\{F\}=\lim_{\max(t_{i+1}-t_i) \rightarrow 0} \ \pi^{-n/2} \prod_{k=1}^{n} (t_k-t_{k-1})^{-1/2}  
\int\limits_{-\infty}^{+\infty} \di x_1 ... \int\limits_{-\infty}^{+\infty} \di x_n \\ F[\{\psi(t_1,...,t_n)(x_1,...,x_n)\}(t)] \, \exp{\lbk - \sum\limits_{k=1}^{n} \frac{(x_k-x_{k-1})^2}{t_k-t_{k-1}} \rbk} \,.
\end{multline}
where Wiener set $F[\{\psi(t_1,...,t_n)(x_1,...,x_n)\}(t)]=F_{t_1, ..., t_n}[f]$ the step functional only depending of $f(t_1), ..., f(t_n)$.

\subsubsection{Developments on Einstein's formula - 1921}
\label{sec_wien_1921}
\noindent

The article on which we focus now was quite particular in Wiener's bibliography. As we already noted, it was hardly ever discussed in the literature, maybe because it was the only one that did not offer novelties from a mathematical point of view, but it is even more interesting to note that Wiener chose a different position compared to his other articles by assuming that the velocities of Brownian particles existed. This article was thus a parenthesis in Wiener's journey, inside which he used a model close to that of Ornstein-Uhlenbeck, which is discussed in \cref{sec_theo_phys_2} .\\

For the first time, Wiener gave a review of the physicists' Brownian movement, which was quite detailed, quoting Einstein and Perrin, explaining the physical origin of the phenomenon in terms of collisions and noting the difference between theoretical and measured velocities due to the `extreme sinuosity of the trajectories'. According to him, Einstein made two assumptions:
\begin{enumerate}[label=(\roman*),align=left,leftmargin=1.75cm]	
	\item the validity of Stokes' law at the scale of Brownian particles, 
	\item the independence of increments over time intervals $\tau$.
\end{enumerate}
His objective was to use the results obtained in his previous article to demonstrate that Einstein's formula on the mean square of displacements (\cref{eq_ecart}) did not in fact require the second hypothesis to be true.

Wiener began by defining a function $ f (ct) $ as the total momentum of the Brownian particle acquired by the collisions with the other molecules only, up to the time $ t $. This was a different definition from the previous article where $ f (t) $ was the position of the particle at time $ t $. The constant $ c $ which depends on the physical parameters must now be noted explicitly because it has an impact on numerical calculations. Wiener could have redefined the normal distribution of \cref{eq_wien_prob} to include $ c $ but he chose to use $ c $ as an argument of the $ f $ function directly, to preserve the form of $ f (t) $ and thus make the results obtained in the previous article usable. According to him the distribution of $ f $ was then given by
\be
\label{eq_proba_impuls_wien}
P \lp a \leq f(ct) \leq b \rp = \frac{1}{\sqrt{\pi ct}}\int_{a}^{b} \exp{\lp-\frac{x^2}{ct}\rp}\di x \,.
\ee
One must note that $a$, $b$ and $x$ have the dimension of a momentum, that is to say $\SI{}{\kilogram\meter\per\second}$, which makes $c$ a quantity expressed in $\SI{}{\kilogram^2\meter^3\per\second\cubed}$.\\

Wiener's starting point was an equation similar to Newton's second law:
\begin{align}
&m \lbk v(t+\di t) - v(t) \rbk =f(ct+c\di t)-f(ct)-6\pi\mu a v(t+ \delta \cdot \di t)\di t &(0 \leq \delta \leq 1) \,.
\end{align}
In this equation the role of collisions was taken into account by $ f (ct + c \di t) -f (ct) $, which appeared as an analogue of Langevin's force $ X $ in \cref{eq_eq_langevin}. The last term accounted for the viscosity and $ \delta $ could vary between 0 and 1, probably to allow the choice of any point of the interval $ [t; t + \di t] $ for the value of the velocity. However this factor did not matter as it would disappear in the next calculation step. Wiener did not make any assumptions about the function $ f $, in particular he did not suppose it to be differentiable, which explained this writing in infinitesimal form. He integrated this equation to obtain
\be
m \lbk v(t)-v_0 \rbk =f(ct)-6\pi\mu a \int_{0}^{t}v(t')\di t' \,,
\ee
where $v_0$ was the initial velocity of the particle. He solved this equation and wrote
\be
mv(t)=mv_0e^{-\beta t}+f(ct)-\beta e^{-\beta t} \int_{0}^{t}f(ct')e^{\beta t'}\di t' \,,
\ee
where we set, in order to simplify the writings,
\be
\label{eq_beta_wiener}
\beta =\frac{6\pi \mu a}{m} \,,
\ee
which is the inverse of the characteristic time $\theta$ defined by Langevin in \cref{eq_temps_carac_lang}.

Since he obtained the velocity, Wiener had to integrate to get the position, then square the expression and take the mean to shape the result in a similar way to that of Einstein. The first step reads
\be
x(t) = \int_{0}^{t} v(t')\di t' = \frac{v_0}{\beta } \lp 1-e^{-\beta t} \rp + \frac{e^{-\beta t}}{mc}\int_{0}^{T}e^{\beta T'/c}f(T')\di T' \,,
\ee
where he made the change of variable $ T = ct $ in the integral. Since $ x (t) $ is a functional of $ f $, the average of $ x^2 (t) $ in the sense of Wiener is therefore written $ A \left \lbrace x^2 (t) \right \rbrace $ (analogue of $ \lambda_x^2 $ in Einstein's notation) and it is an average over the different realizations of the function $ f $ weighted by their probabilities according to \cref{eq_proba_impuls_wien}.
\begin{multline}
A \left\lbrace x^2(t) \right\rbrace = A \left\lbrace \frac{v_0^2}{\beta ^2} \lp 1-e^{-\beta t} \rp^2 + \frac{e^{-2\beta t}}{m^2c^2} \int_{0}^{T} \int_{0}^{T} e^{\beta (T_1+T_2)/c}  f(T_1)f(T_2) \di T_1 \di T_2 \right.  \\
\left. + \frac{2v_0 e^{-\beta t}}{m\beta c} \lp 1-e^{-\beta t} \rp \int_{0}^{T}e^{\beta T'/c} f(T') \di T' \right\rbrace \,.
\end{multline}
It was at this moment that Wiener used \cref{eq_moy_functional def,eq_moy_func_pol} from his previous article, which allowed him to suppress the term in $\left\lbrace f(T) \right\rbrace$, and to explicitly compute the term in $\left\lbrace f(T_1)f(T_2) \right\rbrace$ as
\be
A \left\{ \int_{0}^{T}\int_{0}^{T} f(T_1)f(T_2) H(T_1, T_2) \di T_1 \di T_2 \right\} = \int_{0}^{T}\int_{0}^{T_2} T_1 H(T_1, T_2) \di T_1 \di T_2 \,,
\ee
in the case where $H(T_1, T_2)=H(T_2, T_1)$, which is our case.
After replacement, the formula reads
\be
A \left\lbrace x^2(t) \right\rbrace = \frac{v_0^2}{\beta ^2} \lp 1-e^{-\beta t} \rp^2 + \frac{e^{-2\beta t}}{m^2c^2} \int_{0}^{T} \lbk \int_{0}^{T_2} T_1 e^{\beta (T_1+T_2)/c} \di T_1 \rbk \di T_2 \,.
\ee
Wiener then computed the double integral of the right-hand side and wrote the final result as follows
\be
\label{eq_wien_x2_dvlp}
A \left\lbrace x^2(t) \right\rbrace = \frac{v_0^2}{\beta ^2} \lp 1-e^{-\beta t} \rp^2 + \frac{c}{4m^2\beta ^3}  \lbk 2\beta t - 3 +4e^{-\beta t}-e^{-2\beta t}\rbk \,.
\ee
It appears that the linearity between the average of the squares of displacements and the time comes from the term $ ct / (2m^2 \beta^2) $ in the right-hand side, thus if one wishes to obtain a result similar to Einstein's one, one must make sure that this term is dominant.
Hence, Wiener chose to express the absolute value of the difference between $ A \left \lbrace x^2 (t) \right \rbrace $ and this particular term
\be
\left| \frac{A \left\lbrace x^2(t) \right\rbrace}{t}-\frac{c}{2m^2\beta ^2} \right| =\frac{1}{t} \left|  \frac{v_0^2}{\beta ^2} \lp 1-e^{-\beta t} \rp^2 + \frac{c}{4m^2\beta ^3}   (3-e^{-\beta t})(1-e^{-\beta t}) \right| \,.
\ee 
From this equality, Wiener deduced without justification the following inequality
\be
\left| \frac{A \left\lbrace x^2(t) \right\rbrace}{t}-\frac{c}{2m^2\beta ^2} \right|  \leq  \frac{v_0^2}{\beta ^2} + \frac{3c}{4m^2\beta ^3} \,.
\ee 
It appears that he replaced the exponentials by $0$ to obtain an upper bound of the right hand side, giving that all the terms are positive, but that he forgot to put back the term $1/t$, thus giving a result that was not homogeneous from a physical point of view. For the sake of clarity, we from now on write this missing factor explicitly in the equations. 
 
The last step of Wiener's reasoning consisted in defining the relative deviation as follows 
\be
\label{eq_wiener_ecart_rel}
\frac{\left| A \left\lbrace x^2(t) \right\rbrace / t-c/(2m^2\beta ^2) \right|}{ c/(2m^2\beta ^2)} \leq \frac{1}{t} \lbk \frac{2m^2 v_0^2}{c} + \frac{3}{2\beta } \rbk \,.
\ee 
In order to estimate the right-hand side, Wiener finally referred to Perrin's values on gamboge, without precisely mentioning which ones, and obtained the value $ 10^{-8} $ for the absolute relative error.\\

Let us investigate how Wiener obtained the value $ 10^{-8} $. First, we replace the mathematical notations by their physical counterparts. In Wiener's calculation, the nominal value of $ \lambda_x^2 $ was $ c / 2m^2 \beta^2 $ and in Einstein's theory it was $ k_B T / 3 \pi  \mu a $, hence by equalizing the two quantities and using the definition of $\beta$ given in \cref{eq_beta_wiener} we obtain
\be
c=24 \pi \mu a k_B T \,.
\ee
After replacement, \cref{eq_wiener_ecart_rel} reads
\be
\frac{\left| \lambda_x / t- k_B T/(3 \pi \mu a) \right|}{k_B T/(3 \pi \mu a)} \leq \frac{1}{t} \lbk \frac{m^2 v_0^2}{12 \pi \mu a k_B T} + \frac{m}{4\pi \mu a} \rbk \,.
\ee 
We can suppose the equipartition of energy for the initial velocity, as done by Ornstein in \cref{sec_theo_phys_2}, and thus replace $m v_0^2$ by $k_B T$ (in the one-dimensional case) to obtain
\be
\label{eq_wien_ec_phy}
\frac{\left| \lambda_x / t- k_B T/(3 \pi \mu a) \right|}{k_B T/(3 \pi \mu a)} \leq \frac{1}{t} \frac{m}{3 \pi \mu a} \,.
\ee 
Wiener did not mention the particular set of Perrin's values on gamboge he used but we looked at the ones Perrin gave in \textit{Les Atomes}. On page 182 appears a table of results for several experiments with gamboge, let us choose the line 6 for which the experiment have worked the best, since it was the one that gave the best value for Avogadro's number.
The values are $\mu=\SI{1.25}{\kilogram\per\meter\per\second}$, $a=\SI{0.385}{\micro\meter}$ and $m=\SI{0.29e-12}{\kilogram}$. Replacing these values in \cref{eq_wien_ec_phy} ans choosing\footnote{Indeed, since this factor $t$ did not appear in Wiener's equation at this stage of the calculation, it seemed logical to us to compute this quantity numerically with the value $t=\SI{1}{\second}$. Of course, in reality the error increases as $t$ decreases so the value $ 10^{-8} $ is not a physical limit.} $t=\SI{1}{\second}$, we get $\num{6.4e-8}$, which is in agreement with the order of magnitude calculated by Wiener.
	
We can also note that the right-hand side of \cref{eq_wien_ec_phy} is indeed $2\theta/t$ where $\theta$ is the characteristic time scale defined by Langevin  (\cref{eq_temps_carac_lang}), whose value is $\SI{e-8}{\second}$ according to him. Once again the result is compatible with Wiener's one.\\

Let us analyze how Wiener's mistake impacted the assumption he was trying to prove. He wanted to prove that the hypothesis made by Einstein on the independence of the displacements on disjoint successive time intervals was in fact not necessary, by showing that the result he obtained without making this assumption was so close to Einstein's result that the difference could not be observed experimentally. Therefore, the `slightly different value of  $\lambda_x / t$ for small values of the time than for larger values' observed by experimenters was, according to Wiener, to be traced to the hypothesis (i): a violation of Stokes' law at this scale. However, time $t$ did not appear in the final formula that bounded the absolute relative mistake, so maybe Wiener thought that his value $ 10^{-8} $ held for any time $t$, and thus even for very small values of $t$, hence precisely in the case where discrepancies had been observed. In reality, the upper bound on the relative error is inversely proportional to time $t$, so for very small values of $t$ the upper limit becomes large, making Wiener's conclusion possibly false. This bound reads $2/\beta t$, so if the `small values of time' that invalidated Einstein's result are small but still large enough to ensure $t \gg \beta^{-1}$, then the relative error remains small and Wiener's reasoning still holds. 

Be that as it may, Wiener's calculation had the merit to offer an explicit formula for the upper bound and undoubtedly gave him a legitimacy to later study the idealized Brownian motion, in which he considered that the Gaussian distribution of displacements was valid for all times, as small as desired. This seemed legitimate to him since the physical reality, too complex to model, actually deviated little from the Gaussian result. However, he did not mention this article when he later used his ideal model of Brownian motion.

\subsubsection{Wiener measure - Differential space 1923}
\label{sec_wiener_1923}
\noindent
In 1923, Wiener published \textit{Differential Space}, which would later be cited as the article in which Wiener laid the groundwork for his model of Brownian motion. He used a different approach from the one used so far since he did not base it on Daniell's integral but rather privileged an approach like that of Paul Lévy. Wiener explicitly quoted the exchanges he had with Lévy on the correspondence between his work and Lévy's work, as a starting point for the article.\\

Lévy's theory \citep{levy_cons_1922} can be summarized as follows. We consider functions $ x (t) $ defined on the interval $ [0,1] $. These functions can be approximated by simple functions of order $ n $, which have constant values on each interval $ [0,1 / n] $, ..., $ [(n-1) / n, 1] $. The values taken on each of the intervals are denoted $ x_1 $, ..., $ x_n $. A simple function of order $ n $ is thus represented by a point in a $ n $-dimensional space.
We now consider a particular volume $ V $ in the space of functions defined on $ [0,1] $, let us take the interior volume of the sphere of radius $ R $, defined by
\be
\label{eq_def_sphere_levy}
\int_{0}^{1} x^2(t) \di t =R^2 \,.
\ee
Simple functions of order $ n $ that belong to this domain form a another domain, whose volume $ V_n $ is the volume inside the sphere defined by
\be
\label{eq_def_sphere_simple_levy}
\sum_{i=1}^{n} x_i^2 = n R^2 \,.
\ee
Finally, a functional $ F $ defined and continuous on the volume $ V $, is also continuous and defined on the volumes $ V_n $, on which it has a mean value $ \mu_n $, calculated as a mean of a classical function. Lévy therefore defined the average of $ F $ on $ V $ as the limit of the averages on $ V_n $
\be
\label{eq_def_moy_func_levy}
A \left\{ F \right\} = \lim\limits_{n \rightarrow + \infty} \mu_n \,.
\ee
~\\

Wiener was inspired by this approach when he built his `differential space'. He understood that it was not the successive positions of a Brownian particle that were independent of each other but rather the increments over regular and disjoint time intervals. He then defined the $ n $ increments $ x_n $ from the division of the time axis $ [0,1] $ into $ n $ equal segment, as 
\be
\label{eq_def_diff_wiener}
	\left\{
	\renewcommand{\arraystretch}{1.5}
	\begin{array}{r c l}
		x_1&=&f(\frac{1}{n})-f(0)=f(\frac{1}{n}) \,, \\
		x_2&=&f(\frac{2}{n})-f(\frac{1}{n}) \,, \\
		&... \,, \\
		x_n&=&f(1)-f(\frac{n-1}{n}) \,. \\
	\end{array}
	\right.
\ee	
These $ n $ quantities were independent and had the same statistical weight in the sense that they contributed to the displacement of the particle over equal periods of time, so it seemed natural for Wiener to use Lévy's formulation and to consider the sphere defined by
\be
\label{eq_def_sphere_simple_wiener}
\sum_{i=1}^{n} x_i^2 = R_n^2 \,.
\ee
To test the relevance of this definition, Wiener raised the question of measuring the inner region of the sphere where the position $ f (a) $ was between $ y_1 $ and $ y_2 $, i.e. $ \mu (y_1 \leq f (a) = \sum_ {k = 1}^{na} x_k \leq y_2) $, where the notation $ \mu $ is used for the measure.
In the case where $ R_n = R $ was a constant, Wiener proved that this measure took the value
\be
\label{eq_mes_wien_pos_fin}
\mu(y_1 \leq f(a) \leq y_2)= \frac{1}{\sqrt{2\pi a R^2}} \int_{y_1}^{y_2} \exp{\lp -\frac{u}{2aR^2} \rp} \di u \,.
\ee
This measure was of the same form as the probability given by \cref{eq_wien_prob}, which showed that the Lévy sphere seemed to be an appropriate tool to study Brownian trajectories. Wiener then called the set of functions $ f (t) $, along with the measure $ \mu $ defined as the limit of the measures given in \cref{eq_mes_wien_pos_fin}, the differential space, to reflect the fact that it is the differences that are independent.\\

Wiener returned to his major concern: the definition of the average of a functional. He began once again with the simple case of a functional that depended on a function only by a finite number of its values. He took $ n $ points $ 0 \leq t_1 \ ... t_n \leq 1$ and divided the segment $ [0,1] $ into $ \nu $ parts, all of size $ 1 / \nu $. $ \nu $ must be large enough for only one $ t_i $ value to be included in each interval. He then defined the step function $ f_{\nu} (t) $ which took the values of $ f $ for each value of $ t_i $: $ \forall i, f_{\nu} (t_i) = f (t_i) $. He could then define the differences on the function $ f_{\nu} $: $ x_k = f_{\nu} (k / \nu) -f_{\nu} ((k-1) / \nu) $. Finally, the values $ f (t_1) $, ... $ f (t_n) $ on which depended the functional $ F $ could be expressed in terms of these differences: $ f (t_i) = f_{\nu} (T_i / \nu) = x_1 + x_2 + ... + x_{T_i} $ where he defined $ T_i = \nu t_i $ if $ \nu t_i $ was an integer, or the first higher integer otherwise.

He was thus in the position of applying the formalism he established before, placing himself in the volume defined by the interior of the sphere $ \sum_{i = 1}^{n} x_i^2 = R^2$. The average of the functional within this sphere was properly defined, and if it reached a value when taking the limit, he could speak of the average of this functional on the differential space. The average of $ F $ in the sphere defined above was
\begin{multline}
\label{eq_moy_func_diff_space}
A\{F\}=(2 \pi R^2)^{-n/2} \prod_{k=1}^{n} (t_k-t_{k-1})^{-1/2}  \\
\times \int\limits_{-\infty}^{+\infty} \di y_1 ... \int\limits_{-\infty}^{+\infty} \di y_n \, F[y_1, ..., y_n] \exp{\lbk -\sum\limits_{k=1}^{n} \frac{y_k^2}{2R^2 (t_k-t_{k-1})}\rbk} \,.
\end{multline}
Moreover, Wiener defined the measure of a domain of this sphere by choosing for the functional $ F $ the value 1 if the argument function laid in the considered domain and 0 otherwise. To give an example, he defined the following domain
\be
\label{eq_fenetres_y_diff_space}
\left\{
\renewcommand{\arraystretch}{1.5}
\begin{array}{r c l}	
	
	y_{11} & \leq y_1 \leq& y_{12} \,, \\
	y_{21} & \leq y_2 \leq& y_{22} \,, \\
	&...& \,, \\
	y_{n1} & \leq y_n \leq& y_{n2} \,,
	
\end{array}
\right.
\ee
which was none other than the domain previously defined in \cref{eq_fenetres_x}. Wiener measure of the domain defined by \cref{eq_fenetres_y_diff_space} was
\be
\label{eq_def_mes_diff_pace}
(2 \pi R^2)^{-n/2} \prod_{k=1}^{n} (t_k-t_{k-1})^{-1/2}  \int\limits_{y_{11}}^{y_{12}} \di y_1 ... \int\limits_{y_{n1}}^{y_{n2}} \di y_n \, \exp{\lbk -\sum\limits_{k=1}^{n} \frac{y_k^2}{2R^2 (t_k-t_{k-1})} \rbk} \,.
\ee

This long article was full of other rich ideas but we chose to focus on the one that was a first step towards the non-differentiability of Brownian trajectories.

Wiener reversed the previous perspective, in which he was interested in the distribution of $ f (a) = \sum_{k = 1}^{na} x_k $ with the constraint $ \sum_{i = 1}^{n} x_i^2 = R^2 $. Inversely, he raised the question of the distribution of $ \sum_{i = 1}^{n} x_i^2 = \sum_{i = 1}^{n} \lbk f (i / n) -f ((i- 1) / n) \rbk^2 $ assuming that the increments $ f (t_2) -f (t_1) $ were independent and had a Gaussian distribution. He demonstrated that the quantity $ \sum_ {i = 1}^{n} x_i^2 $ only deviated from $ D $ with a small probability, where $D$ was the physical diffusion coefficient.
\begin{align}
\label{eq_proba_var_born_brown}
&\forall \varepsilon>0,\ \forall \delta>0,& P\lp \left| \lim\limits_{n \rightarrow + \infty} \sum_{1}^{n} \lbk f \lp \frac{k}{n} \rp -f \lp \frac{k-1}{n} \rp \rbk^2 - D \right| > \delta \rp < \varepsilon \,,
\end{align}
Next, he considered the particular case of a continuous functions $f$, of limited total variation.
\begin{align}
\sum_{k=1}^{n} \lbk f \lp \frac{k}{n} \rp -f \lp \frac{k-1}{n} \rp \rbk^2 &\leq \max_{k} \left| f \lp \frac{k}{n} \rp -f \lp \frac{k-1}{n} \rp \right| ~\ \sum_{k=1}^{n} \left| f \lp \frac{k}{n} \rp -f \lp \frac{k-1}{n} \rp \right| \nonumber \\
& \leq T(f) \cdot \max_{k} \left| f \lp \frac{k}{n} \rp -f \lp \frac{k-1}{n} \rp \right| \label{eq_coef_vat_born} \,,
\end{align}
where $ T (f) $ was the total variation of the function $ f $ over $ [0,1] $. We recall that the total variation of a function $ f $ is a possible measure of its arc length and can be defined as $ T (f) = \sup_{\mathcal {P}} \sum_{k = 1}^{n} \left| f \lp k / n \rp -f \lp (k-1) / n \rp \right| $ where $ \mathcal {P} $ denotes the set of partitions (in the modern sense, not that of $ P_K $) of $ [0,1] $.

When making $ n $ tend to infinity in \cref{eq_coef_vat_born}, the upper bound in the right-hand side went to $ 0 $. Since Wiener showed that in the case of Brownian trajectories the quantity $ \sum_{i = 1}^{n} \lbk f (i / n) -f ((i-1) / n) \rbk^2 $ did not tend to $ 0$ but was close to $D$, he therefore deduced that it was infinitely unlikely for Brownian trajectories to be of limited total variation and therefore infinitely unlikely that they have a bounded derivative.\\

Wiener called the coefficient $ \lim_{n \to \infty} \sum_{i = 1}^{n} \lbk f (i / n) -f ((i-1) / n) \rbk^2 $ `a kind of coefficient of non-differentiability of the function $ f $'\footnote{It should be noted that in the article it was the quantity $ \lim_{n \to \infty} \sum_{i = 1}^{n} \lbk f (i / n) -f ((i-1 ) / n)^2 \rbk $ that Wiener called the coefficient of non-differentiability but it was a printing error, as there were several in the article.} because it measured a degree of non-differentiability equal to its deviation from $ 0 $.

\subsubsection{Lebesgue measure and stochastic processes - 1930}
\label{sec_wiener_1932}
\noindent
The next advances in Brownian motion modelling came later, in the 1930s, with a thesis on harmonic analysis published in 1930 by Wiener \citep{wiener_generalized_1930} and with the collaboration with mathematicians Paley and Zygmund in 1932-1933, which culminated to a fundamental article in 1933 \citep{paley_notes_1933}.
In his 1930 article, Wiener gave a new formulation of Brownian motion, in terms of Lebesgue measure, and it was this formulation that allowed Brownian motion to be considered as a stochastic process in the 1933 article. This led to the proof of the non-differentiability of Brownian trajectories, which ends our history of mathematical Brownian motion.
From this time on, Wiener's work became harder to follow for non-mathematicians, so we try to make the most of the arguments without going into technical details.\\

From 1924, Wiener had been interested in harmonic analysis, which was the study of the properties of Fourier series expansions. He wrote a series of articles on harmonic analysis, including an important memoir in 1930 \citep{wiener_generalized_1930}.
In this thesis, Wiener studied general questions about the probability distributions of the frequency or energy spectra of time-dependent functions. It was therefore a science at the border between statistics and harmonic analysis.
Chapter 13 of this thesis named `Spectra and Integration in Functional Space' opened with a discussion of the coexistence of these two sciences in physics. According to Wiener, the question had been mainly raised in optics, when it was necessary to obtain statistical behaviours of the superposition of electromagnetic waves, as for their amplitude or their intensity. Rayleigh had contributed the most to this study using a theory based on statistical harmonic analysis. Wiener's goal was to show that a better and more rigorous approach than Rayleigh's one could be thought out. The following general question arose: if a resonator received a chaotic sequence of input pulses, it would reemit the input spectrum but amplify certain contributions and reduce others; how then to study the statistics of this output? To approach the problem it was necessary to have a clear vision of what `chaotic' meant and the simplest chaotic system for Wiener was Brownian motion. He aimed to rigorously define the integration on the space of continuous functions defined on the segment $ [0,1] $, then to extend his results to the set of continuous functions defined on $ \mathbb{R} $. Indeed, a function defined on $ \mathbb{R} $ could be expanded in Fourier integral and allowed the use of all the results developed by Wiener in the previous chapters of his memoir; while the functions on $ [0,1] $ could be studied as Fourier series, which was a slightly different framework. 

Wiener recalled that he had dealt with the problem of integration on functions space in the past, with Daniell's theory of integration, but that his new approach was different: his goal was to achieve a mapping between the functions of $ C_0[0,1] $ and the segment $ [0,1] $. If such a mapping was established, then each path $ f(t) \in C_0[0,1] $ would be uniquely represented by a point of the segment $ [0,1] $ and vice versa. Eventually, this correspondence made it possible to integrate on the $ [0,1] $ segment with Lebesgue measure rather than on the set $ C_0 [0,1] $ with Wiener measure, which was more convenient for the rest of the memoir. We do not discuss how this new representation contributed to the rest the memoir but we rather outline the construction of the mapping, which would be taken for granted in the 1933 article \citep{paley_notes_1933}.

The idea of the construction was quite simple and relied entirely on the one made in the article \citep{wiener_average_1922}, discussed in the \cref{sec_wiener_1920}. For the sake of clarity, we use the definitions of the divisions $ I_n $, intervals $ i(I_n) $, and weights $ w[i(I_n)] $, given by \cref{eq_fenetres_x,eq_wien_proba_intervalles}. To avoid confusion and preserve the coherence of the notations we keep the name interval only for the intervals $ i(I_n) $ of a division, and refer to the interval $ [0,1] $ as segment. His idea was to first establish the mapping between a division $ I_n $ and the segment $ [0,1] $ by placing the intervals of this division on sub-parts of the segment whose lengths were equal to the weights of the intervals.

To illustrate this, let us take the easiest case: $ n = 1 $. According to the definition of the divisions, there are two time values $ t_1 = 1/2 $ and $ t_2 = 1 $. For these values, the functions $ f $ must pass through the windows $ [x_{11}, x_{12}] $ and $ [x_{21}, x_{22}] $. The possible values for $ k_h $ are $ 0 $ and $ 1 $, so there are 4 intervals in this division, defined by the choices of $] - \infty, 0] $ or $ [0, + \infty [$ independently for $ [x_{11}, x_{12}] $ and for $ [x_{21}, x_{22}] $. The 4 intervals are obviously of equal weight, worth 1/4. Thus we cut the segment $ [0,1] $ into four sub-segments of length 1/4 and we assign to each interval one of the four sub-segments of $ [0,1] $, thus matching the weight of the interval with the length of the sub-segment that is associated with it.

We repeat the operation with $ n = 2 $ and the $ 4^4 = 256 $ intervals of $ I_2 $. If, for a set of values of the parameters $ k_h $, which defines an interval $ i_j (I_2) $, the interval $ i_j (I_2) $ is found to be included in an interval of the type $ i (I_1) $ (and the uniqueness of the latter is ensured by the hypothesis (i) of the partitions, discussed in \cref{sec_wiener_1920}), then we place the sub-segment corresponding to $ i_j (I_2) $, the length of which is defined by the weight of this interval, inside the sub-segment corresponding to $ i (I_1) $.

Wiener repeated the operation with $ n \rightarrow + \infty $. The intervals of the successive divisions thus fitted in infinitely until reaching sub-segments of zero length. Thus each point of the segment was determined by an infinite sequence of intervals whose weights tend to 0, defining a unique trajectory of zero probability. Wiener therefore completed the construction of the mapping between the points of the $ [0,1] $ segment and the functions of $ C_0[0,1] $. \\

In 1933, Wiener, Paley and Zygmund published an article together, that aimed at establishing a correspondence between Wiener's theory on the one hand, and Paley and Zygmund's theory on the other hand. Indeed, the two theories dealt with the introduction of randomness in analysis but with different approaches, which could be unified. Paley and Zygmund's work focused on the study of random-coefficients series, initiated by Borel, and that of Wiener dealt with the study of random trajectories such as those of Brownian motion, whose study he reduced to a problem of Lebesgue integration, as we saw previously. The aim was therefore to establish similar theorems to those obtained by Paley and Zygmund for the random functions studied by Wiener.

We present here two aspects of this paper: the formalism used to describe Wiener's random functions and the Theorem VII, which established the almost surely non-differentiability of Brownian trajectories.

The authors proposed to represent
\begin{quote}
one of the components of the displacement of the moving particle in the Wiener theory by $\chi(\alpha,t)$, where $t$ is the time and $\alpha$ the parameter on which Lebesgue integration is performed for the purpose of averaging over all functions and determining probabilities. Wiener's $\chi(\alpha,t)$ is then, as he shows, a continuous function of $t$ for almost all values of $\alpha$, defined for $0\leq \alpha \leq 1$, $0 \leq t \leq 1$ and vanishing for $t=0$. 
\end{quote}
This definition may seem surprising at first glance, but it was nevertheless the way to define the stochastic processes later (for example in \citep{doob_wieners_1966}) and it still is today. $ \alpha $ is a number between $0$ and $1$, which according to the mapping established by Wiener in the previous article represents a function of $ C_0 [0,1] $, which we also note $ \alpha $. We observe a change of perspective, rather than denoting $ \alpha (t) $ a component of the trajectory of a Brownian particle, we use a $ \chi $ function with two variables, taking as argument both the time $ t $ and the trajectory $ \alpha $. An enlightening interpretation of this quantity was given in \citep{doob_wieners_1966}.

$ \chi (\alpha, t) $ must be understood as the value of the $ \alpha $ trajectory at time $ t $, i.e. $ \chi (\alpha, t) = \alpha ( t) $. Then $ \chi (\alpha, \cdot) $, seen as a function of $ t $, is a trajectory, or a realization, of the stochastic process.
With this formalism, Joseph Léo Doob summarized the properties Wiener had established on Brownian motion as
\begin{enumerate}[label=(\roman*),align=left,leftmargin=1.75cm]
	\item Its increments are Gaussian: $\chi(\cdot,t)$ is a random variable on $C_0[0,1]$, following the normal law $\lp \chi(\cdot,t_2) - \chi(\cdot,t_1) \rp \sim \mathcal{N}(0,\sigma^2 |t_2-t_1|)$,
	\item its increments on disjoint time intervals are independent: $\forall\ 0\leq t_0 \leq...\leq t_n \leq 1$, random variables $\chi(\cdot,t_1) - \chi(\cdot,t_0)$, ..., $\chi(\cdot,t_n) - \chi(\cdot,t_{n-1})$ are independent.
\end{enumerate}
~\\
Let us go back to the 1933 article. The authors obtained, once more, the formula of the average of a functional when this one depended only on a finite number of values given at times $ t_0 =0 \leq t_1 \leq ...\leq t_n \leq 1$, but with the $\chi$ formalism 
\begin{align}
\label{eq_moy_func_proc_stoc}
A\{F\}&=\int_{0}^{1} F\lbk \chi(\alpha,t_1), \chi(\alpha,t_2) - \chi(\alpha,t_1), ..., \chi(\alpha,t_n) - \chi(\alpha,t_{n-1}) \rbk \di\alpha \nonumber \\
&= \pi^{-n/2} \prod_{k=1}^{n} (t_k-t_{k-1})^{-1/2} \int\limits_{-\infty}^{+\infty} \di x_1 ... \int\limits_{-\infty}^{+\infty} \di x_n \, F[x_1, ..., x_n] \, \exp{\lbk -\sum\limits_{k=1}^{n} \frac{x_k^2}{t_k-t_{k-1}} \rbk} \,.
\end{align}

At the end of the article, the authors proved for the first time the property of non-differentiability of mathematical Brownian trajectories in a rigorous way. The result can be written as follows
\begin{theorem}
	The values of $\alpha$ for which there exists a $t$ such that 
	
	\begin{align*}
	&\lim\limits_{\varepsilon \rightarrow 0}~\ \frac{\chi(\alpha,t+\varepsilon)-\chi(\alpha,t)}{\varepsilon^{\lambda}} < \infty &(\lambda > 1/2) \,,
	\end{align*}
	
	form a set of zero measure
\end{theorem}
In other words, the probability of choosing a Brownian trajectory from the set $ C_0 [0,1] $, for which there exists at least one time point at which the trajectory is differentiable, is zero.
We note that the proof for $ \lambda = 1 $ would have been enough, since this is the definition of differentiability, but the theorem actually expressed a stronger property.

It was also in this article that the authors introduced the first stochastic integral, formally noted as 
\be
\label{eq_int_stoch_wien} 
\int_{0}^{1} f(x,t)c(t)\di_t\chi(\alpha,t) \,,
\ee
where one had to make sense of `$ \di_t \chi (\alpha, t) $', since they showed that the function $ \chi $ did not have a derivative. This problem is part of what is called stochastic calculus, which was strongly developed by Doob in the years that followed and which allowed to connect Langevin's and Einstein's approaches by proving a formal equality between Langevin's white noise and this Brownian motion's `derivative'. We do not discuss this point nor Doob's stochastic calculation because it is a vast subject going far beyond our scope, but it is historically important enough to be mentioned.\\

Let us close this section with Wiener's own words
\begin{quote}
	To my surprise and delight I found that the Brownian motion as thus conceived had a formal theory of high degree of perfection and elegance. Under this theory I was able to confirm the conjecture of Perrin and to show that, except for a set of cases of probability $0$, all the Brownian motions were continuous non-differentiable curves. (\cite{wiener_i_1956}, p.39)
\end{quote}

\section{Physicists' Brownian motion after 1908}
\label{sec_theo_phys_2}
\noindent
After Perrin's experiments, the attention of physicists largely turned to the experimental application of the physical theory of Brownian motion. The formulas derived by Einstein, Smoluchowski and Langevin were used to measure fundamental quantities such as Avogadro's number, appearing explicitly in \cref{eq_ecart}, or the elementary charge of the electron. For example, a series of articles on the measurement of the elementary charge using Brownian motion, written by Jean Perrin, Louis de Broglie, or Felix Ehrenhaft, was published in the \textit{Comptes rendus de l'académie des sciences de Paris} from 1908.  

However, other physicists worked on enhancing the first theories of Brownian motion, in order to give them more generality. This movement came mainly from the Dutch school, with the two authors we are going to study: Leonard Ornstein and George Uhlenbeck, but also with other figures who played a role, such as Gertruida de Haas-Lorentz or Willem Rhijnvis van Wijk. 

Despite the importance of Ornstein-Uhlenbeck process in the field of stochastic processes, their construction has not been analyzed in details, therefore in this section, we aim to examine the theory developed by these physicists, starting with Ornstein's 1917 article, and ending in 1934 with an article by Ornstein and van Wijk. We wish to answer the questions: what are the gaps they wanted to fill in former theories? and how their contributions were intimately linked to the notion of velocity and its very existence at short time scales? by studying the primary literature.
This story ran parallel to Wiener's one, and their comparison and interactions are discussed in \cref{sec_comparison}.

\subsection{Leonard Ornstein and George Uhlenbeck's works}
\label{sec:OU}
\noindent
Leonard Ornstein was a Dutch physicist who contributed a lot to statistical physics and stochastic processes, and gave his name to the Ornstein-Uhlenbeck process, which we discuss later.
He studied physics and obtained his thesis at Leiden University under the direction of Hendrick Lorentz in 1908. He then became professor of mathematical physics at Groningen University in 1909 and professor of theoretical physics at Utrecht University, succeeding Peter Debye, in 1914. He remained in Utrecht, but his interest turned to experimental physics when he took over the post of lab head of experimental physics in Utrecht in 1925. His laboratory gained a great renown under his direction thanks to the measurements of luminous intensity which he obtained by revolutionary methods of photometry \Citep{mason_leonard_1945}.

The first article on Brownian movement communicated by Ornstein in 1917, but published only in 1919 \citep{ornstein_brownian_1919}, cited Gertruida de Haas-Lorentz's work as the starting point. De Haas-Lorentz was Hendrick Lorentz's daughter, under whose direction she obtained her thesis in physics from Leiden University in 1912. She was, in her thesis, the first to use calculations of fluctuations on electrons seen as Brownian particles. Her thesis was cited in numerous articles on Brownian motion by Dutch physicists, for the theory that was exposed but also for her historical review of the work on Brownian motion up to 1912. Unfortunately it has not yet been translated into English.
In his article, Ornstein gave a generalization of Einstein's \cref{eq_ecart}, valid for all times, as well as a similar relation for speed increments. He also developed a Fokker-Planck relation on probability distributions, similar to that on position given by Einstein (\cref{eq_diffus_pos}), but for the distribution of velocities, and valid for all times.

Ornstein wrote a second article on Brownian movement in 1918, also published in 1919, with the physicist Frederik Zernike \citep{ornstein_theory_1919}. We do not analyse this article because its subject is out of the scope of our study. \\

The following biographical elements about George Uhlenbeck's can be found and supplemented by reading \citep{cohen_george_1990, dresden_george_1989}. Uhlenbeck was also a Dutch physicist, twenty years younger than Ornstein. He entered Leiden University in 1919, and studied physics with Paul Ehrenfest, under whose direction he obtained his thesis in 1927. Indeed, in 1912 Lorentz decided to end his teaching activities at the university to devote more time to research, and proposed to Einstein to take his place. The latter refused because he had just accepted a job at ETH Zurich, and Ehrenfest took over from Lorentz in Leiden.
After his thesis, Uhlenbeck held the position of professor of physics at Michigan University in Ann Arbor from 1927 to 1935, then became professor of theoretical physics at Utrecht University in 1935, replacing Hans Kramers, who himself inherited the chair of theoretical physics in Leiden, left empty after Ehrenfest's suicide. In 1939, Uhlenbeck returned to  Michigan University and remained there until 1960. Between 1935 and 1939, both Ornstein and Uhlenbeck were at Utrecht University.
Uhlenbeck marked the history of physics thanks to several major works including the introduction of the concept of spin in 1925 with Samuel Goudsmit, also Ehrenfest's PhD student, and his work on stochastic processes including the introduction of Ornstein- Uhlenbeck process and the first appearance of the master equation in an article on cosmic rays in 1940.

Uhlenbeck's interest in Brownian movement came from his reading of Ornstein's works, however his first article on the subject was not co-written with him. Indeed, Uhlenbeck was working on the interpretation of quantum physics and he wrote his first article on Brownian movement in 1929 with Goudsmit, with whom he had already worked on the spin a few years earlier, while the two young doctors were professors at Michigan University.
This paper discussed the rotational Brownian motion of small suspended mirrors used in a 1927 experiment by Walther Gerlach; we do not talk about it. 

Uhlenbeck and Ornstein's paths finally crossed in their common article in 1930 \Citep{uhlenbeck_theory_1930}, which represented the birth of Ornstein-Uhlenbeck process. In this article, the authors attempted to compute moments of order greater than 2, for displacements and for velocity, in order to obtain complete probability distributions. Indeed, in his 1917 article, Ornstein had only calculated the moments of order 1 and 2, which was enough to generalize Einstein's results, but which was insufficient to find the whole probability distribution, valid at all times. They also tried to get the Fokker-Planck equation associated with the displacement distribution, valid for any time, but failed.
Finally, in 1934 this difficulty was overcome when Ornstein published an article with van Wijk, a former student of his who obtained his thesis under Ornstein's supervision in 1930 at Utrecht university. In this article \citep{ornstein_derivation_1934} they derived the Fokker-Planck equation for the distribution of displacements. The long time limit of this equation restored the diffusion equation, and the short time limit gave an equation accounting for the hydrodynamic aspects of the motion. 

We analyze the 1917 article in a first part, and the articles of 1930 and 1934, which completed the lines suggested in 1917, in a second part. These papers also marked major advances in the theorization of Brownian motion in the presence of external forces, such as harmonic potential or coupled potentials, but we do not tackle these points because they reduce to the study of Langevin's equation with an additional term and bring nothing to the general understanding of the theory.

\subsubsection{Einstein's formula at short times - Ornstein 1917}
\label{sec_orn_1917}
\noindent
Ornstein exposed the objectives of his article as follows: from the relation used by de Haas-Lorentz in her 1912 thesis, and used to derive the relation connecting the mean of the squares of the deviations to time (\cref{eq_ecart}), the probability function of Brownian motion can be determined. In addition, Ornstein claimed a new method for the calculation of averages, and to obtain the distribution in velocity, in addition to that in position.

What was de Haas-Lorentz's relation? This was actually Langevin's equation (\cref{eq_eq_langevin}), but Ornstein did not cite him at any time. It is interesting to note that in his next article \citep{ornstein_theory_1919}, he quoted Langevin twice in the expression `Langevin-Einstein Formula' referring to the preceding article, without further details on this formula.
We learn a little more in the 1934 article \citep{ornstein_derivation_1934} in which the authors spoke of Langevin equation in these terms `This equation is generally known as the equation of Einstein. According to F. Zernike, Langevin used it before Einstein.' For Ornstein, it appears that De Haas-Lorentz's relation and that of Langevin-Einstein are the same, so we decided to name it Langevin's equation, as it was the case until now.

It is true that Ornstein presented a new way of calculating the averages, based on the properties imposed on Langevin's stochastic force $ X $, and deduced the first two moments of the position and velocity distributions, valid at any time, but complete distributions were obtained only in 1930 with Uhlenbeck. \\

Ornstein started with the relation
\be
\label{eq_orn_lang}
\frac{\di v}{\di t}=-\beta  v + F \,,
\ee
where $\beta =6\pi \mu a/m$ (it is the same definition as Wiener's one in his article \Citep{wiener_average_1921}) and where $F$ was Langevin's stochastic force $X$ divided by mass $m$. By formally integrating this equation, Ornstein got
\be
\label{eq_orn_v}
v=v_0 e^{-\beta  t} + e^{-\beta t} \int_{0}^{t} e^{\beta t'}F(t')\di t' \,,
\ee
where $v_0$ was the initial velocity of the particle. He then averaged at a given time over many particles having the same initial velocity, using the fact that $\langle F \rangle=0$ because collisions are irregular and uncorrelated, and got
\be
\label{eq_orn_v_fin}
\langle v \rangle = v_0 e^{-\beta  t} \,.
\ee
Thus, velocity decreases exponentially because of viscous drag.

Ornstein repeated the operation by first squaring \cref{eq_orn_v} before averaging to obtain $\langle v^2 \rangle$, which made a term in $\langle F(t')F(t'') \rangle$ appear
\be
\label{eq_orn_v2}
\langle v^2 \rangle=v_0^2 e^{-2\beta  t} + e^{-2\beta t} \int_{0}^{t} \int_{0}^{t} e^{\beta (t'+t'')} \langle F(t')F(t'') \rangle \di t'\di t'' \,.
\ee 
Ornstein was the first to propose a quantitative treatment for the term $ \langle F (t ') F (t'') \rangle $. According to him, there were correlations between collisions only for very short periods, i.e. when $t'$ and $t''$ were very close, otherwise $F (t ')$ and $F (t'')$ would be independent and therefore the average of their product would be null. He defined $\psi $ as $ t '' = t '+ \psi $, hence $ \langle F (t ') F (t'') \rangle $ was non-null only for values of $ \psi $ very close to $ 0 $. From this statement, he used a series of approximations.
He first replaced $ t '' $ by $ t '$ in the exponential term and was then able to factor the double integral into a product of two integrals as
\be
\int_{0}^{t} dt' e^{2\beta t'} \lbk \int_{0}^{t} \langle F(t')F(t'+\psi) \rangle \di\psi \rbk \,.
\ee
Since the term in the second integral is non-null only for values of $\psi$ close to $0$, he extended the integration bounds of this integral to $] - \infty, + \infty [$, without changing the result significantly. He finally defined $ \gamma $, a constant which depended on the parameters of the problem and which would be computed later, as
\be
\label{eq_orn_def_gamma}
\int_{-\infty}^{+\infty} \langle F(t')F(t'+\psi) \rangle \di\psi=\gamma \,.
\ee
After replacing this in the main expression, Ornstein obtained
\be
\langle v^2 \rangle = v_0^2 e^{-2\beta  t} + \gamma \frac{1-e^{-2\beta  t}}{2\beta } \,.
\ee
It remained only to explain the constant $ \gamma $ according to the parameters of the problem, for that Ornstein used the equipartition of energy once the equilibrium reached, i.e. in the long time limit
\be
\label{eq_orn_val_gamma}
\lim\limits_{t \rightarrow \infty} \langle v^2(t) \rangle = \frac{k_B T}{m} \ \ \Rightarrow \ \ \gamma = \frac{2 \beta  k_B T}{m} \,.
\ee
Finally, Ornstein obtained an expression for $\langle v^2 \rangle$, analogous to the one for the displacements (\cref{eq_ecart}), but true at all times
\be
\label{eq_orn_v2_final}
\langle v^2 \rangle = v_0^2 e^{-2\beta  t} + k_B T \frac{1-e^{-2\beta  t}}{m} \,.
\ee
He did not discuss this expression, or test his behaviour in the regime of small times, but went directly to the calculations on displacements. By integrating again \cref{eq_orn_v}, he made the displacement $ s $ appear, defined as $s=x-x_0$ where $x$ is the position and $x_0$ the initial position
\be
\label{eq_orn_eqdiff_x}
\beta  s = v_0-v+\int_{0}^{t}F(t')\di t' \,.
\ee
Following the same logic as for the velocity, he squared this expression, and used the results of \cref{eq_orn_v_fin,eq_orn_v2_final,eq_orn_def_gamma} to obtain
\be
\label{eq_orn_full_v0}
\beta ^2 \langle s^2 \rangle = v_0^2 \lp 1- 2e^{-\beta  t} +e^{-2\beta  t} \rp +\frac{\gamma}{2\beta } \lp -3 -e^{-2\beta  t}+4e^{-\beta  t} \rp + \gamma t \,.
\ee
This equality was exactly the one Wiener would get (\cref{eq_wien_x2_dvlp}) four years later, independently. It should be noted that contrary to Einstein's formula, the initial velocity $ v_0 $ appeared explicitly. In the long time limit, Ornstein's formula gave back Einstein's formula, and cancelled the dependency on the initial velocity
\be
\langle s^2(t) \rangle \underset{t \rightarrow \infty}{\sim} \frac{\gamma}{\beta ^2} t = \frac{k_B T}{3 \pi \mu a} t \,.
\ee
To obtain a result which was completely comparable to that of Einstein, even at short times, it remained to average \cref{eq_orn_full_v0} over the different initial velocities $v_0$, by replacing the value of $ v_0^2 $ by the one predicted by the equipartition of energy: $ \langle v_0^2 \rangle = k_B T / m $. We from now on write $ \langle \cdot \rangle_{v_0} $ for the statistical average on all particles and on all initial velocities. Therefore the equation read
\be 
\label{eq_orn_x2}
\beta ^2 \langle s^2 \rangle_{v_0} = \frac{\gamma}{\beta }\lp \beta  t -1 + e^{-\beta  t}\rp \,.
\ee
It appears that Einstein's formula is valid when the first term in the parenthesis dominates, i.e. when $ \beta t> 1 $ or equivalently when $ t> m / 6 \pi \mu a $, which is consistent with the range of validity defined by Langevin.

Ornstein did not discuss in this article the short time limit of this formula, but he did in the 1930 article with Uhlenbeck. \\

In the last part of his article, Ornstein established the Fokker-Planck equation whose solution was the velocity distribution. The aim was to give an expression, similar to \cref{eq_diffus_pos}, for the distribution $ f (v, v_0, t) $, in the case where $ f (v, v_0, t) \di v $ represented the number of particles having a velocity between $ v $ and $ v + \di v $ at time $ t $, and an initial velocity $ v_0 $. If the first part of the article was strongly based on Langevin's approach, Ornstein followed from there the logic used by Einstein to obtain the diffusion equation, discussed in the \cref {sec_einstein}.

Ornstein integrated \cref{eq_orn_lang} over a very short time $\tau$, and got
\be
v=u(1-\beta \tau)+ \xi \,,
\ee
where $u$ is the velocity at the beginning of the time interval, and where he defined
\be
 \xi=\int_{0}^{\tau}F(t)\di t \,.
\ee
He used the statistical properties of $ F $ to define a probability distribution $ \phi_{\tau} (\xi) $ so that $ \langle \xi \rangle = \int_{0}^{\infty} \xi \phi_{ \tau} (\xi) \di \xi = 0 $ because $ \langle F \rangle = 0 $, and $ \langle \xi^2 \rangle = \int_{0}^{\infty} \xi^2 \phi_{\tau } (\xi) \di \xi = \gamma \tau $, by definition of $ \gamma $ (\cref{eq_orn_def_gamma}). Thus, following Einstein's idea, Ornstein wrote that the number of particles $ f (v, v_0, t + \tau) \di v $ having a velocity between $ v $ and $ v + \di v $ at time $ t + \tau $ was equal to the number of particles $ f (u, v_0, t) \di u \phi_{\tau} (\xi) \di \xi $ having a velocity between $ u $ and $ u + \di u $ at time $ t $ and having received a velocity increment of value $ \xi $ due to the force $ F $, integrated over all the values of $ \xi $ 
\begin{align}
f(v,v_0,t+\tau)\di v&=\int_{-\infty}^{+\infty} f(u,v_0,t)\di u \ \phi_{\tau}( \xi) \di \xi \nonumber \\
&=(1+\beta \tau)\di v \int_{-\infty}^{+\infty} f(u,v_0,t) \ \phi_{\tau}( \xi) \di \xi \,.
\end{align}
Ornstein took the Taylor expansion of $f$ to the first order in $\tau$ and to the second order in $u$ about $v$. Using the properties of $\phi_{\tau}$ he got
\be
\frac{\partial f}{\partial t}=\beta  f + \beta v (1+\beta \tau) \frac{\partial f}{\partial v} + (1+\beta \tau)\frac{\gamma}{2} \frac{\partial^2 f}{\partial v^2} \,.
\ee
The last step was to make $ \tau $ tend toward $ 0 $. This was one of the main differences with Einstein's derivation for which a physical hypothesis set the acceptable value of $ \tau $, whereas in our present case, no independence was imposed and therefore the time $ \tau $ can be taken arbitrarily small. In this limit, we have
\be
\label{eq_orn_fp_vit}
\frac{\partial f}{\partial t}=\beta  \frac{\partial}{\partial v} \lp v \cdot f \rp + \frac{\gamma}{2} \frac{\partial^2 f}{\partial v^2} \,.
\ee
This equation is the diffusion equation for velocity, analogous to \cref{eq_diffus_pos} for position, but valid at all times. One can note, as Ornstein did, that the parameter $ \gamma $, defined by \cref{eq_orn_def_gamma}, and whose value is given by \cref{eq_orn_val_gamma}, plays the role of the diffusion coefficient in velocity space, and is therefore analogous to the diffusion coefficient $ D $ in position space. Combining \cref{eq_coef_dif,eq_beta_wiener,eq_orn_val_gamma} we find the relation 
\be
\label{eq_diff_coef_rel}
\gamma = 2 \beta^2 D \,.
\ee
Ornstein did not fully solve the problem in this article, as he only solved this equation in steady state, that is to say with the left-hand side null. Doing so, he found the Maxwell-Boltzmann distribution for the steady state velocity distribution.\\

Let us take a few moments to list fundamental differences between the speed distribution and the position distribution. Although the physical phenomenon is the same: a Brownian particle moves and one chooses to study either its position or its speed, Fokker-Planck equations whose solutions are the two distributions in question differ by a convective term. Such a term can also appear in the Fokker-Planck equation for position if one considers the case of a particle subjected to a force, which we do not study here. The two phenomena are said to be diffusive, in a broad sense in the case of velocity because of the additional convective term, however the diffusion coefficients of the two phenomena are different. Finally, the velocity distribution reaches a steady state: the Maxwell-Boltzmann distribution, while the position distribution is explicitly time-dependent at any time and does not approach a stable distribution; it is a Gaussian law that flattens and expands indefinitely. Even if the physical experiment is the same, owing to the differences listed above, it makes sense to give to the two stochastic processes different names. When one is interested in the diffusion of the velocity of a Brownian particle, one speaks of a Ornstein-Uhlenbeck process, and when one is interested in the diffusion of the position of the Brownian particle, one speaks of a Wiener process.

\subsubsection{Probability laws at short times - Ornstein \& Uhlenbeck 1930}
\noindent
In 1930, Ornstein and Uhlenbeck wrote an article together, in continuity with Ornstein's 1917 article. Back then, he had generalized Einstein's formula, giving the average square of displacements as a function of time, for any time and had given a similar formula for the average square of velocities. He had also obtained the Fokker-Planck equation associated with the velocity diffusion process. The two authors' aim was therefore to finish the construction by determining the complete distributions for displacements and velocities, valid at all times, and to obtain the Fokker-Planck equations of the two processes, with a general method. In this article we can recognize Uhlenbeck's style, known to be very clear and orderly \Citep{dresden_george_1989}: he began by giving a detailed reminder of all the results obtained so far, in the form of a review article, then established clearly the points he aimed to deal with, and in the end explained the limits of his work and the directions to be pursued.

For each of the objectives they set, the authors used a different method: they determined the probability distributions by the method of moments, that is by calculating all the moments $ \{\langle q^k \rangle \}_{k \in \mathbb {N}} $, which fully characterize the distribution; and obtained the Fokker-Planck equations by a general method similar to that of Ornstein in 1917. 

Moments 1 and 2 had already been computed by Ornstein, so the authors recalled the results and proposed an interpretation of \cref{eq_orn_x2} in the short time limit
\begin{align}
\langle s^2 \rangle_{v_0}& \underset{t \rightarrow 0}{\sim} \frac{k_B T}{m}t^2 \nonumber \\
& \underset{t \rightarrow 0}{\sim} \langle v_0^2 \rangle t^2 \,.
\end{align}
This relation was fundamental, as it was the main difference with Einstein's formula (\cref{eq_ecart}). Indeed, Einstein's prediction took the form $\langle s^2 \rangle \propto t$, leading to the impossibility to define a velocity; whereas Ornstein and Uhlenbeck obtained a law in the form $\langle s^2 \rangle_{v_0} \underset{t \rightarrow 0}{\propto} t^2$, which allowed the interpretation of a uniform displacement at velocity $\sqrt{\langle v_0^2 \rangle}$. This law is independent of the viscosity of the medium, which can be understood with the interpretation of the viscosity as the result of the collisions of the suspended particle with the medium particles. Yet, in the short-time limit, this law describes the movement of the particle between two collisions.

For higher moments, they started by the case of velocity. The two authors announced that the Gaussian distribution held for all times, but for the modified variable $\mathcal{V}=v-v_0 e^{-\beta t}=v- \langle v \rangle$. If $\mathcal{V}$ has a Gaussian distribution, then its moments must satisfy the following properties
\be
\label{eq_orn_cond_gaus}
\left\{
\renewcommand{\arraystretch}{1.5}
\begin{array}{c c l}	
	
	\langle \mathcal{V}^{2n+1} \rangle &=& 0 \,, \\
	\langle \mathcal{V}^{2n} \rangle &=& 1 \cdot 3 \cdot 5 \cdot ... \cdot (2n-1) \ \langle \mathcal{V}^{2} \rangle^n \,.
	
\end{array}
\right.
\ee
Thus, one just needs to compute the moments of $\mathcal{V}$ and to check that they satisfy the above relations, to be sure that $\mathcal{V}$ follows a Gaussian law. From the first two moments already computed by Ornstein (\cref{eq_orn_v_fin,eq_orn_v2_final}), they knew that
\be
\label{eq_orn_v_moments_1_2}
\left\{
\renewcommand{\arraystretch}{1.5}
\begin{array}{c c l}	
	
	\langle \mathcal{V} \rangle &=& 0 \,, \\
	\langle \mathcal{V}^{2} \rangle &=& \frac{k_B T}{m} \lp 1- e^{-2\beta t} \rp \,.
	
\end{array}
\right.
\ee
The method they used to compute the other moments was essentially the same as for the first two: they used the statistical properties of $ F $ to calculate the averages and then used the equipartition of energy at equilibrium to determine the constants. Doing so, they obtained the moments $ \langle \mathcal{V}^3 \rangle $ and $ \langle \mathcal{V}^4 \rangle $, and left the general case to the reader. Let us analyse these two calculations. 
They first put \cref{eq_orn_v} to the desired power, then took the average and used the assumptions on the distribution of $ F $. We do not detail this step which was technical and required the introduction of new assumptions, such as the one made in \cref{eq_orn_def_gamma} for the average of a product of two terms $ F $, but for products of over three terms $ F $, which gave rise to new constants $ C_i $, similar to $ \gamma $ but for larger products. 
There was no additional physical ingredient in this step, the principle was still to consider that the function $ F $ was very sharp around $ 0 $ and that, as a consequence, products of type $ F (t_1). .. F (t_n) $ were non-zero only close to the domain $ t_1 = ... = t_n $.
We rather focus on the second step in which the physical reasoning appeared.

Ornstein and Uhlenbeck obtained
\be
\langle \mathcal{V}^{3} \rangle = \frac{C_1}{\beta } \lp 1- e^{-3\beta t} \rp \,.
\ee
To determine the constant, it must be noted that $\mathcal{V}$ and $v$ have the same distribution when $t \rightarrow + \infty$. Since they supposed that steady state velocities $v$ followed a Maxwell-Boltzmann distribution by hypothesis, then
\begin{align}
\lim\limits_{t \rightarrow + \infty} \langle \mathcal{V}^{3} \rangle &= \lim\limits_{t \rightarrow + \infty} \langle v^{3} \rangle \nonumber \\
&=0 \,,
\end{align}
so $C_1=0$ and $\langle \mathcal{V}^{3} \rangle =0$, which was required by \cref{eq_orn_cond_gaus}.

Let us look at their calculation for the fourth moment now
\be
\langle \mathcal{V}^{4} \rangle = \frac{3 \gamma^2}{4\beta ^2} \lp 1- e^{-2\beta t} \rp^2 + \frac{C_2}{2\beta } \lp 1- e^{-4\beta t} \rp \,.
\ee 
Once again, they determined the constant $C_2$ using the Gaussian distribution of velocities $v$ once the steady state was reached
\begin{align}
\lim\limits_{t \rightarrow + \infty} \langle \mathcal{V}^{4} \rangle &= \lim\limits_{t \rightarrow + \infty} \langle v^{4} \rangle \nonumber \\
&=3 \lim\limits_{t \rightarrow + \infty} \langle v^{2} \rangle^2 \nonumber \\
&=\frac{3 \gamma^2}{4 \beta ^2} \,.
\end{align}
It followed that $C_2=0$ and $\langle \mathcal{V}^{4} \rangle=3 \langle \mathcal{V}^{2} \rangle^2$.

The moments of higher orders were computed in the same way and eventually the two authors deduced that $\mathcal{V}$ followed a Gaussian law, as announced, which took the form
\be
\label{eq_orn_distrib_v}
f(v,v_0,t)=\sqrt{\frac{m}{2 \pi k_B T \lp 1- e^{-2\beta t} \rp}} \, \exp{\lbk -\frac{m}{2 k_B T} \frac{\lp v -v_0 e^{-\beta t} \rp^2}{1- e^{-2\beta t}} \rbk} \,.
\ee
This transformed into Maxwell-Boltzmann distribution in the long-time limit, and into a Dirac distribution at initial velocity $v_0$ for the $t=0$ limit, as expected. \\

The problem at the level of displacements is to find the probability for a particle that started at time $t=0$ at position $x_0$ with velocity $v_0$ to lie between positions $x$ and $x+\di{x}$ at time $t$. The computations were performed in a similar manner and made constants $C_i$, already determined, appear. Ornstein and Uhlenbeck showed that the Gaussian law for displacements also held for any time, but for the modified variable $\mathcal{S}=s-v_0 \lp 1- e^{-\beta t} \rp /\beta = s - \langle s \rangle$, and could be written
\be
\label{eq_orn_distrib_x}
f(x,x_0,t)=\sqrt{\frac{m\beta ^2}{2 \pi k_B T \lp 2\beta t -3 +4 e^{-\beta t}- e^{-2\beta t} \rp}} \, \exp{\lbk - \frac{m\beta ^2}{2 k_B T} \frac{\lbk x-x_0 -\frac{v_0}{\beta } \lp 1- e^{-\beta t} \rp \rbk^2}{2\beta t -3 +4 e^{-\beta t}- e^{-2\beta t}}\rbk} \,.
\ee
Interestingly, the velocity distribution depends only on the initial velocity $v_0$, whereas the displacement distribution depends on both $v_0$ and the initial position $x_0$. In the short time limit, this distribution converges to a Dirac function at $x_0$, as expected. In the long time limit however, at the numerator of the fraction inside the exponential remains the term $x-x_0 -v_0/\beta $, whereas only $x-x_0$ appears in Einstein's formula \cref{eq_distrib_pos}. The explanation given in the following article \citep{ornstein_derivation_1934} was that  `For time intervals large in comparison with [$1/\beta$], we may neglect [$\frac{v_0}{\beta } \lp 1- e^{-\beta t} \rp $] compared with $x-x_0$; since $|x-x_0|$ will not remain finite for large values of t' whereas $v_0/\beta $ remains constant. In fact this argument is a bit incorrect because position $x$ and time $t$ in the probability distribution have to be considered as independent variables taking any values, but not supposing a dependency of $x$ on $t$. The result is correct though, and the difference with Einstein's formula is accounted for by the fact that Ornstein and Uhlenbeck considered the case of a particle moving with initial velocity $v_0$, unlike Einstein. Their probability distribution should then be written rigorously $f(x,x_0,v_0,t)$. \\

In this article, the authors also offered a general method to obtain Fokker-Planck equations, which they applied to the particular cases of the velocity distribution, to rediscover Ornstein's 1917 result, and of the displacement distribution. The derivation followed the same scheme as Einstein's one, detailed in \cref{sec_einstein}, and as Ornstein's one, presented in \cref{sec_orn_1917}, but with some significant differences in the hypotheses, which we examine now.

Let $f(q,q_0,t)$ be the distribution of a variable $q$, with initial value $q_0$ at time $t=0$. The associated Fokker-Planck equation is the partial differential equation whose $f$ is solution. Let us consider that during a time $\Delta t$ the variable $q$ changes by an amount $\Delta q$, with a probability distribution $\phi(\Delta q,q,t)$, depending on the value $q$ at the beginning of the time interval $\Delta t$, but which we suppose to be independent of the initial value $q_0$. We write with a prime symbol the value of $q$ after the increment: $q'=q+ \Delta q$. Following the same reasoning as before, they got 
\be
\label{eq_orn_ulh_fp1}
f(q', q_0, t+\Delta t) = \int_{-\infty}^{+\infty} f(q'-\Delta q, q_0, t)\phi(\Delta q, q'-\Delta q, t) \di (\Delta q) \,.
\ee
Like their predecessors, they went on with a Taylor expansion in $\Delta t$ for the left-hand side and in $\Delta q$ for the right-hand side. Both expansions were for the moment complete, that is to say they did not choose to cut the Taylor series at a particular order, and consequently they had an infinity of terms. The right-hand side expansion gave rise to the apparition of all the moments of $\Delta q$ : $\left\lbrace \langle \Delta q^k \rangle \right\rbrace_{k \in \mathbb{N}}$. 
The equation contained two infinitesimal quantities: $\Delta t$ and $\Delta q$, and in order to get a useful equation for physicists, they had to neglect terms from a particular order and therefore to choose where to stop the expansion for both infinitesimal quantities coherently. For that, the authors defined two functions
\begin{empheq}[left = \empheqlbrace]{align}
\lim\limits_{\Delta t \rightarrow 0} \frac{\langle \Delta q \rangle}{\Delta t}&=g_1(q,t) \label{eq_orn_uhl_def_f1} \,, \\
\lim\limits_{\Delta t \rightarrow 0} \frac{\langle \Delta q^2 \rangle}{\Delta t}&=g_2(q,t) \label{eq_orn_uhl_def_f2} \,, 
\end{empheq}
and assumed that for higher orders
\be
\label{eq_orn_uhl_cond_f3}
\lim\limits_{\Delta t \rightarrow 0} \frac{\langle \Delta q^k \rangle}{\Delta t}=0 \hspace{40pt} k \geq 3 \,.
\ee
This hypothesis was in fact equivalent to keeping terms of order 1 in $\Delta t$ and terms of order 2 in $\Delta q$, in the case where $\Delta t$ tended to $0$. Indeed, dividing the expression that contained the full expansions by $\Delta t$ and letting $\Delta t$ tend to $0$ made all terms of order greater than or equal to 2 in $\Delta t$ null, and all terms of order greater than or equal to 3 in $\Delta q$ null as well, according to \cref{eq_orn_uhl_cond_f3}. This was the very choice that Einstein made to obtain the diffusion equation. 

Using these definitions and letting $\Delta t$ tend toward $0$, they finally obtained
\be
\label{eq_orn_uhl_fp_general}
\frac{\partial f}{\partial t} = \frac{g_2}{2} \frac{\partial^2 f}{\partial q^2}+ \lp \frac{\partial g_2}{\partial q} - g_1 \rp \frac{\partial f}{\partial q} + \lp \frac 1 2 \frac{\partial^2 g_2}{\partial q^2} - \frac{\partial g_1}{\partial q} \rp f \,.
\ee
This is a general Fokker-Planck equation for any system that satisfies the hypothesis given by \cref{eq_orn_uhl_cond_f3} and for which increments $\Delta q$ are independent of the initial value $q_0$. In order to use it for their problem, the authors needed to determine the functions $g_1$ and $g_2$ and to make sure that \cref{eq_orn_uhl_cond_f3} was satisfied. \\

From all the previous developments, it was easy to check that $ g_1 (v, t) = -v \beta $, $ g_2 (v, t) = \gamma $ for the case of velocity, and that the condition of \cref{eq_orn_uhl_cond_f3} was satisfied. By replacing these values in \cref{eq_orn_uhl_fp_general}, they found the Fokker-Planck equation associated with the diffusion of velocities, already obtained by Ornstein in 1917 (\cref{eq_orn_fp_vit}).

A difficulty arose when it came to applying this method to displacements. They needed to compute $ \langle \Delta x \rangle $ and $ \langle \Delta x^2 \rangle $ to get $ g_1 $ and $ g_2 $, but when averaging \cref{eq_orn_eqdiff_x} they obtained
\begin{align}
- \beta \langle \Delta x \rangle &= \langle v' \rangle - \langle v \rangle \nonumber \\
&=v_0 e^{-\beta t} \lp e^{\beta \Delta t} -1 \rp \,,
\end{align}
thus in the $\Delta t \rightarrow 0$ limit, it became
\be
\label{eq_orn_uhl_delta_x}
\langle \Delta x \rangle =v_0 e^{-\beta t} \Delta t \,.
\ee
Following the same calculations by putting first the equation to the square, they got $\langle \Delta x^2 \rangle \propto \Delta t^2$ and so $g_2$ was null. Then Fokker-Planck equation read
\be
\label{eq_orn_uhl_fp_hydro}
\frac{\partial f}{\partial t} = -v_0 e^{-\beta t} \frac{\partial f}{\partial x} \,.
\ee
This equation could not be the general Fokker-Planck equation of the problem since it did not transform into Einstein's diffusion equation in the $t \gg \beta^{-1}$ limit, as it should have.

Ornstein and Uhlenbeck noted this difficulty and attributed it to the hypothesis they made when deriving the Fokker-Planck equation in the general case (\cref{eq_orn_uhl_fp_general}) and which seemed defective in this case. The hypothesis in question was the independence of the increments $ \Delta x $ with respect to the initial values $ x_0 $ and $ v_0 $. Indeed, we can see in the calculation that this hypothesis was not verified since the average of the increments $ \langle \Delta x \rangle $ depended explicitly on $ v_0 $ according to \cref {eq_orn_uhl_delta_x}

This difficulty was overcome only four years later, by the joint efforts of Ornstein and Van Wijk in their 1934 article, in which they obtained the true Fokker-Planck equation without using the previously mentioned hypothesis, therefore in a more general case, which allowed its application to the case of the displacements. In reality, it was not the relaxation of this hypothesis that made it possible to obtain the announced Fokker-Planck equation. Following the reasoning of the two authors, we however explicitly note the dependence on the initial condition $ q_0 $ in the probability distribution of the increments $ \phi (\Delta q, q, q_0, t) $. The derivation performed in the previous article is still valid because the $ q_0 $ dependency did not play a direct role. The only modifications to take into account are the dependencies in $ q_0 $ of all the quantities derived from $ \phi (\Delta q, q, q_0, t) $. This is the case of the moments $ \langle \Delta q^k \rangle $ of this distribution and consequently of the functions $ g_1 $ and $ g_2 $ which we thus explicitly note $ g_1 (q, q_0, t) $ and $ g_2 (q, q_0, t) $. The general equation (\cref {eq_orn_uhl_fp_general}) was therefore still valid but Ornstein and van Wijk admitted they could not deduce the equation for the case of displacements from this reasoning. Their next idea was quite surprising and could appear as a mathematical sleight of hand. They assumed that the functions $ g_1 $ and $ g_2 $ were only functions of time, and called them $ n (t) $ and $ m (t) $ respectively. This assumption was obviously not verified a priori but allowed them to eliminate the derivatives of these functions with respect to $ q $ in \cref{eq_orn_uhl_fp_general}, which gave
\be
\label{eq_orn_wijk_fp_simpl}
\frac{\partial f}{\partial t} = \frac{m(t)}{2} \frac{\partial^2 f}{\partial q^2} - n(t) \frac{\partial f}{\partial q} \,.
\ee
By defining
\begin{empheq}[left = \empheqlbrace]{align}
N(t)&=\int_{0}^{t}n(t')dt' \,, \\
M(t)&=2\int_{0}^{t}m(t')dt' \,,
\end{empheq}
the exact solution read
\be
f(q,q_0,t)=\frac{1}{\sqrt{\pi M(t)}} \, \exp{\lbk -\frac{\lp q - N(t) \rp^2}{M(t)} \rbk} \,.
\ee
Hence, $f(q,q_0,t)$ followed a Gaussian distribution for the variable $q$, centred on $N(t)$ and of variance $M(t)/2$
\begin{empheq}[left = \empheqlbrace]{align}
N(t) &= \langle q \rangle \,, \\
M(t) &=2 \lbk \langle q^2 \rangle - \langle q \rangle^2 \rbk \,.
\end{empheq}
Functions $n(t)$ and $m(t)$ were given by the derivatives of the above results
\begin{empheq}[left = \empheqlbrace]{align}
n(t)&=\frac{d}{dt}\langle q \rangle \,, \\
m(t)&=\frac{d}{dt} \lbk \langle q^2 \rangle - \langle q \rangle^2 \rbk \,.
\end{empheq}

To finish the calculation, they used the value of the moments $\langle q \rangle$ and $\langle q^2 \rangle$ given by the displacement distribution (\cref{eq_orn_distrib_x})
\begin{empheq}[left = \empheqlbrace]{align}
n(t)&=v_0 e^{-\beta t} \,,\\
m(t)&=\frac{2 k_B T}{m \beta} \lp 1-e^{-\beta t} \rp^2 \,.
\end{empheq}
Replacing these values in the simplified Fokker-Planck equation (\cref{eq_orn_wijk_fp_simpl}), they had
\be
\label{eq_orn_wijk_fp_x_final}
\frac{\partial f}{\partial t} = \lp 1-e^{-\beta t} \rp^2 \frac{k_B T}{m \beta} \frac{\partial^2 f}{\partial x^2}- v_0 e^{-\beta t} \frac{\partial f}{\partial x} \,.
\ee
This reasoning may seem illegitimate because of the unrealistic assumption made but it ultimately gave the expected result, as clear-mindedly explained by the authors
\begin{quote}
The mode of reasoning by which [\cref{eq_orn_wijk_fp_x_final}] has been obtained from [\cref{eq_orn_uhl_fp_general}] may seem anything but stringent, because [$g_1$] and [$g_2$] will certainly depend on $x$ too in the case in question but however this may be.. [\cref{eq_orn_wijk_fp_x_final}] is a differential equation possessing [\cref{eq_orn_distrib_x}] as its fundamental solution and transforming into the diffusion equation for $t= \infty$. On the other hand, if $t=0$, it gives the equation of hydrodynamics [\cref{eq_orn_uhl_fp_hydro}] which describes the flow of an ensemble of particles. Thus, the whole range of $t$ values is covered by [\cref{eq_orn_wijk_fp_x_final}] and it is very instructive to see from [\cref{eq_orn_wijk_fp_x_final}] in what way the equation of motion for an ensemble of particles all having the same position $x_o$ and velocity $v_0$ at $t = 0$ is transformed into the diffusion equation by the action of the unsystematic impulses of the surrounding liquid. (\cite{ornstein_derivation_1934}, p. 253-254)
\end{quote}

\section{Mathematical and physical Brownian motion as parallel theories}
\label{sec_comparison}
\noindent
We have studied in the previous sections Ornstein-Uhlenbeck's theory and Wiener's theory of Brownian motion, that are both based on former theories developed by Einstein, Smoluchowski and Langevin in the 1900s.
Since they have been developed in parallel over the same time period but in different directions and pursuing different goals, we raise the questions of the comparison between the constructions of the two theories in a first part, then we ask if there were any interactions between the two theories in a second part, and finally we aim to shed light on potential differences between the possibilities offered by the two theoretical frameworks in a third part.

\subsection{Comparison between the philosophies of both theories}

Regarding the analyses conducted in the previous sections, it appears clearly that the existence of Brownian velocity is at the heart of the divergence between Wiener's and Ornstein-Uhlenbeck's works.

Indeed, Ornstein-Uhlenbeck theory assumed by construction the existence of the velocity of Brownian particles and was built in order to complete Einstein's theory which was lacking results at short time scales, where the velocity became ill-defined. On the other hand, 
Wiener extended Einstein's result to any time, as small as desired, by building an idealized Brownian motion. His construction led to the non differentiability of trajectories, that is to say the non existence of the velocity, caused by the Gaussian distribution for displacements at all times.

This difference in the primary assumptions to the theories can be explained by the diverging goals of the authors. Wiener was interested in the study of the properties of the curves representing idealized Brownian trajectories, even if he was aware that these trajectories were not physical trajectories, but only a `surrogate'. He aimed and succeeded at showing that these trajectories were described by continuous functions nowhere differentiable, as first guessed by Perrin. 
On the opposite, the Dutch movement focused on giving a meaning to Einstein's theory at short time scale, in the sense of correcting the formulas in order for the results to be physically acceptable. They thus obtained a well-defined velocity corresponding to the physical intuition of a particle traveling in straight line between two collisions. 

We can also mention that Wiener took Einstein's results as a starting point for his construction, without mentioning Langevin's approach, while the Dutch worked both from Langevin's equation, to obtain the different moments of physical quantities, and from Einstein's approach on probability distributions, called the Fokker-Planck method\footnote{As we mentioned in \cref{sec:langevin}, although Einstein wrote a Fokker-Planck equation in 1905, which consisted of starting from an integral relation on a probability distribution to obtain a partial differential equation on the same distribution, it was named after the physicists Adrian Fokker and Max Planck who obtained it in 1913, in a more general framework than that of Einstein.}, to obtain the partial differential equation of the problem. 

The only exception to the differences mentioned in the preceding paragraph is the article \citep{wiener_average_1921}, discussed in \cref{sec_wien_1921}, in which Wiener did not follow the current development of his theory at that time, but rather started from an equation similar to that of Langevin, supposing the existence of the velocity of Brownian particles and turned to what happens at Einstein's time scale $ \tau $. This article, unique in Wiener's bibliography, can be compared very directly to some of Ornstein's results, as we saw in previous sections.

\subsection{Did mathematical Brownian motion influence physical theories?}

One may wonder whether Ornstein and Uhlenbeck were aware of Wiener's work, and if so, what justified for them the use of a model where velocity existed. To provide some answers, we must look at two later articles. Firstly, we examine Joseph Léo Doob's 1942 article  \Citep{doob_brownian_1942}, in which he proposed the first rigorous mathematical theory of Ornstein-Uhlenbeck process, which in particular dealt with the existence of the time derivative of the velocity within Orstein-Uhlenbeck's contruction. Secondly, we quote Uhlenbeck and Ming Chen Wang's 1945 review article \citep{uhlenbeck_theory_1945}, in which they mentioned Wiener's and Doob's works, which offers some insight on the way Uhlenbeck was aware of mathematical works and used it or not for physical purposes.

\subsubsection{Doob's development on Ornstein-Uhlenbeck theory}
\label{sec_doob}

The years 1930-1940 were marked by an important development of modern probability theory, that is to say probability based on measure theory coming from analysis; in particular with Lévy's, Kolmogorov's and Doob's works. This modern theory gave birth to stochastic processes, on the boundary between mathematics and theoretical physics. A new vocabulary appeared for new mathematical objects.

Doob was one of the first mathematicians to formulate a theory of stochastic processes with continuous parameters (e.g. time) at a time when most probabilists were not fond of measure theory \citep{getoor_j._2009}. Doob wrote his first article on the subject in 1937, but it is his fundamental article published in 1942 that we are particularly interested in. In this article appeared the first rigorous mathematical theory of Ornstein-Uhlenbeck process, expressed in modern terms stemming from the theory of stochastic processes. Doob's stated objectives were: to use modern probability methods to analyse Ornstein-Uhlenbeck distributions and to give the absolute probability distributions, as opposed to the conditional distributions that Ornstein and Uhlenbeck offered. These distributions are conditional because they depend on the initial values $ q_0 $ of the concerned parameters, they are therefore said to be distributions of $ q $ knowing $ q_0 $.

Thanks to his modern analysis, Doob demonstrated that the velocity $ v $, appearing in Ornstein-Uhlenbeck process, did not admit a derivative, that is to say that Brownian particles had no finite acceleration. His argument was quite similar to the one for the non-differentiability of displacements in the case of Gaussian increments, he showed that 
\begin{align}
\langle \lbk v(t+s) - v(s) \rbk^2 \rangle &= 2 \ \frac{k_B T}{m} \lp 1 - e^{-\beta t} \rp \nonumber \\
& \underset{t \rightarrow 0}{\sim} 2 \ \frac{k_B T}{m} \beta t \,.
\end{align}
Consequently, $\sqrt{\langle \lbk v(t+s) - v(s) \rbk^2 \rangle }$ was of the order of magnitude of $\sqrt{t}$ at short times, and therefore the velocity was non-differentiable. 

Following this statement, he sought a proper writing of Langevin equation in order to avoid writing $\di v/ \di t$, which was not defined. It was on this occasion that Doob introduced a new differential writing, which gave rise to stochastic integrals, such as those initiated by Wiener, of which we spoke at the end of \cref{sec_wiener_1932}. He proposed the following new formulation for Langevin equation
\be
\label{eq_stoch_doob}
\di v(t)=-\beta v(t)\di t + \di B(t) \,,
\ee
where $B(t)$ was a stochastic noise which must be specified. The above equation was simply Langevin equation multiplied by $\di t$, in which Doob set $\di B(t)=F(t)\di t$. He showed that $ B $ had the same properties as a Wiener process and thus that the Gaussian white noise $ F $ was formally the derivative of a Wiener process, although the latter did not admit a derivative in the strict sense.

Doob then worked to make sense of \cref{eq_stoch_doob} and the different terms that composed it. According to him, we must understand this equation as equivalent to any equation of the form
\be
\label{eq_stoch_doob_int}
\int_{a}^{b} f(t) \di v(t)=-\beta \int_{a}^{b} f(t)v(t)\di t + \int_{a}^{b} f(t)\di B(t) \,,
\ee
for all real $a$ and $b$ and for all continuous function $f$. 
The first two integrals were classical integrals while the third one was a stochastic integral.

Doob solved \cref{eq_stoch_doob_int} by choosing $f(t)=\exp{(\beta t)}$ and setting $v_0=0$
\be
\label{eq_doob_int}
v(t)=e^{-\beta t} \int_{0}^{t} e^{\beta t'} \di B(t') \,.
\ee
This result is the equivalent of Ornstein's \cref{eq_orn_v}, but was derived using an other expression of Langevin equation, that gave rise to the stochastic integral in the right hand side. \\

To close this section, we wish to convey Jean-Pierre Kahane's insightful analysis on the way Doob's model completes the circle of Brownian motion theories. Kahane noted in his lecture on Paul Langevin at the \textit{Bibliothèque Nationale de France} \citep{kahane_paul_2014} that the physicists' Brownian motion presupposed the existence of the velocity while that of Wiener demonstrated that velocity did not exist. According to him, they were therefore incompatible, but complementary. He further explained that the two echo each other by construction. Indeed, one can start from the Langevin equation, which relies on a well-defined velocity of the particle, to obtain Einstein's relation on the mean of the squares of displacements in function of the time. Then Wiener started from this last formula to construct an idealized theory of Brownian motion, in which velocity did not exist. With Doob we complete the circle in the other direction because Langevin equation, that controls the evolution of the velocity of a particle, can be written formally with the help of the mathematicians' Brownian motion, that is the Wiener process. The solution to that Langevin equation can also be expressed using Wiener process. Indeed, Kahane explained in \citep{kahane_essai_1998} that the solution to \cref{eq_doob_int} can be written in terms of a Wiener process $W_1$:
\be
v(t)= \frac{e^{-\beta t}}{2 \beta} W_1(e^{2\beta t}) \,.
\ee

\subsubsection{Ornstein-Uhlenbeck and Wiener's disagreement on the status of velocity}

In 1945, Uhlenbeck wrote a new article \citep{uhlenbeck_theory_1945} on Brownian movement in collaboration with Ming Chen Wang, a colleague of Michigan University, with whom he co-wrote eleven articles on kinetic-theory problems. The authors proposed to redo a review article on the different Brownian movement theories, with the new vocabulary developed in the years 1930-1940. It was in this publication that Ornstein-Uhlenbeck process was expressed as a stochastic process for the first time.

It was also in this 1945 article, twelve years after Wiener's article in which he demonstrated the non-differentiability of Brownian trajectories in the framework of the Wiener process, and three years after Doob's article in which he proved the non-differentiability of the velocity of Brownian particles in the framework of the Ornstein-Uhlenbeck process, that Uhlenbeck referred to these two works for the first time.
\begin{quote}
	The authors are aware of the fact that in the mathematical literature (especially in papers by N. Wiener, J. L. Doob, and others; cf. for instance [\citep{doob_brownian_1942}], also for further references) the notion of a random (or stochastic) process has been defined in a much more refined way. This allows for instance to determine in certain cases the probability that the random function $y(t)$ is of bounded variation, or continuous or differentiable, etc. However, it seems to us that these investigations have not helped in the solution of problems of direct physical interest, and we will, therefore, not try to give an account of them. (\cite{uhlenbeck_theory_1945}, p. 324, footnote 9)	
\end{quote}
We thus see that the authors knew the results of the mathematicians, moreover they quoted some of Wiener's articles explicitly in the rest of their article. They believed, however, that the models constructed by mathematicians were not of physical interest and did not contribute to the analysis of experimental results. Indeed, even though the non-differentiability of trajectories had given rise to many developments in mathematics, it seemed hardly acceptable from a physical point of view, although evoked by Perrin in the first place. Despite the extreme sinuosity of the trajectories, there was a scale below which the particle moved in straight line at a constant velocity between two collisions, which was accounted for by Ornstein and Uhlenbeck's extension to short times of the first physical theories. The contributions for physics of the mathematical models of Brownian motion rather lay in the development of mathematical tools that would be used by physicists, and even chemists and biologists, to deal with other stochastic processes appearing in natural sciences.\\

\subsection{Is Ornstein-Uhlenbeck theory better than Wiener's one?}

According to Uhlenbeck, mathematicians' Brownian motion, based on Einstein-Smoluchowski theory, contributed nothing to the understanding of physical experiments. We shall go beyond that statement and ask what the theories are able to describe and predict, and in particular if one theory fails to address a topic while the other does not. 

In \citep{nelson_dynamical_1967}, Edward Nelson made a clear statement about the similarity of the predictions of both theories in ordinary cases whereas they could differ in precise cases:
\begin{quote}	
	For ordinary Brownian motion (e.g., carmine particles in water) the predictions of the Ornstein-Uhlenbeck theory are numerically indistinguishable from those of the Einstein-Smoluchowski theory. However, the Ornstein-Uhlenbeck theory is a truly dynamical theory and represents great progress in the understanding of Brownian motion. Also, as we shall see later (Chapter 10), there is a Brownian motion where the Einstein-Smoluchowski theory breaks down completely and the Ornstein-Uhlenbeck theory is successful. (\cite{nelson_dynamical_1967}, p. 45)	
\end{quote}
Let us analyze his arguments behind each affirmation. 

To show that predictions were indistinguishable, he used the same method as Wiener in his article \citep{wiener_average_1921}, discussed in \cref{sec_wien_1921}, that is, he computed the absolute relative error between the variance of displacements computed with Ornstein-Uhlenbeck theory and with Einstein's theory. According to him\footnote{This is straightforward from \cref{eq_orn_full_v0} or from \cref{eq_wien_x2_dvlp}.}, the variance of displacements in Ornstein-Uhlenbeck's theory was given by
\be
\langle x^2 \rangle - \langle x \rangle^2 = 2Dt+\frac{D}{\beta} \lp -3 + 4 e^{-\beta t} -e^{-2 \beta t} \rp \,,
\ee
where the second term in the right-hand side was the deviation from the value $ 2Dt $ predicted by Einstein's theory. He estimated the value $ 3 \cdot 10^{- 8} $ for the upper bound of the absolute relative error, using the values $ \beta^{- 1} = \SI{e-8}{\second} $ and $ t = \SI{1/2}{\second} $. He found the same order of magnitude as the one computed by Wiener, and since this error was not experimentally measurable as long as we remain in the domain where $ t \geq \beta^{-1} $, Einstein's theory seemed to be a very good approximation of Ornstein-Uhlenbeck's exact theory.

Nevertheless, there was a case where the predictions of the exact theory and of the approximation differed significantly: Brownian motion in the presence of an external force, which we have not discussed in this article.
It seems important to us to briefly mention Nelson's argument because it gives legitimacy to Ornstein-Uhlenbeck theory. Indeed, their theory was a great theoretical progress but without significant predictions on experimental grounds, if one considers only classical Brownian motion. 
Nelson considered a Brownian particle in a harmonic potential of pulsation $ \omega $, and wrote Langevin equation\footnote{This Langevin equation written in differential form follows the formalism developed by Doob in \cref{eq_stoch_doob}, which we discussed in \cref{sec_doob}.} in the coupled form
\begin{empheq}[left = \empheqlbrace]{align}
\di x(t)&=v(t)\di t \,, \\
\di v(t)&=-\omega^2x(t)\di t-\beta v(t) \di t +\di B(t) \,,
\end{empheq}
where $B$ was a Wiener process of variance\footnote{The variance of the Wiener process was set to the value $2\beta^2 D$, which is what Ornstein called $\gamma$, according to \Cref{eq_diff_coef_rel}. This is therefore in agreement with the primary definition of $\gamma$, given by \cref{eq_orn_def_gamma} (variance of $F$).} $2\beta^2 D$. There were then three cases:
\begin{enumerate}[label=(\roman*),align=left,leftmargin=1.75cm]
	\item $\beta > 2 \omega$ : over-damped, the viscous friction force is stronger than the harmonic force,
	\item $\beta < 2 \omega$ : under-damped, the viscous friction force is weaker than the harmonic force,
	\item $\beta = 2 \omega$ : critically damped.
\end{enumerate}
This experiment was carried out by Eugen Kappler in 1931 in the three cases. Brownian particles were small suspended mirrors that were impacted by surrounding gas molecules. This was indeed a one-dimensional Brownian motion, described by the angle that the mirrors made with their equilibrium positions around their axis, while they were subjected to a torsion force of harmonic form.
In case (i) the result was expected to be very close from the Brownian motion of a free particle, and therefore Einstein's approximation was expected to be valid. 
The plot of the angle versus time in case (i) was very similar to a Wiener trajectory (noisy, random, without a regular pattern), although it never deviated much from its median position because of the torsion force. This was the curve of a Markov process, satisfyingly approximated by a Wiener process, which is itself Markovian.
In case (ii) however, because of the dominance of the harmonic potential, the trajectory was much smoother and approached a sinusoid, which could not be accounted for by a Markov process. In this case, Einstein-Smoluchowski's theory, or equivalently Wiener processes, failed to account for this experimental behaviour.

\section{Conclusion}
\noindent
In this article, we reviewed the history of Brownian motion from Einstein's first article in 1905 to Ornstein's last article in 1934. We investigated a transition period that has often been overlooked, when Brownian motion became an object of interest for mathematicians. Through the concept of velocity, we questioned the motivations of the actors during this period, and explored an aspect of the relationship between physics and mathematics by testing the links between theories. Within this study, we explicated the status of velocity in different Brownian motion theories, we tackled the birth of stochastic processes and the different ways to introduce randomness into physics, and we presented Wiener's results in a physicist-friendly way. 

Regarding the first physical theories of Brownian motion, we showed that Einstein and Smoluchowski developed quite similar models in which the notion of displacement was preferred to that of velocity, which was not properly defined. They introduced stochasticity in their theories by considering quantities coming from probability theory, namely probability distributions or average quantities. On the contrary, Langevin considered an equation in which velocity appeared, and brought randomness into this equation thanks to an additional stochastic force. Those two ways of thinking the stochastic aspect of a problem are still employed today and named Fokker-Planck equation and Langevin equation. As long as the time $t$ was not too small, all three theories agreed on the fundamental formula $\langle x^2 \rangle \propto t$, which is a consequence of the Gaussian distribution of displacements.

We investigated Wiener's background in order to identify the reasons of his interest for Brownian motion, until then a playground for physicists. Wiener's taste for physics and his awareness of Brownian motion are partially explained by the readings his Cambridge's professors suggested. In addition, he found in Brownian motion a good testing ground for his ideas on integration theory, which he developed in the light of Perrin's quotation on the similarity between Brownian trajectories and functions without tangent. He built the first mathematical theory of Brownian motion, taking Einstein's theory as a starting point but knowingly working on an idealized version of it, in which the Gaussian distribution discussed above holds at time. This choice led him to a proof of the non-differentiability of Brownian trajectories, that is, the non-existence of Brownian velocities, thus echoing Perrin's intuition. 

Ornstein, later joined by Uhlenbeck, took another road alongside Wiener's work. Whereas Wiener decided to extend Einstein's Gaussian distribution to short times, Ornstein wanted to complement Einstein's results with a proper treatment of the short time limit that respected physical intuition. He chose to work with the velocity, following Langevin's early calculations. For these two reasons, Wiener's and Ornstein-Uhlenbeck's theories were conflicting and yet complementary, as confirmed by the modern stochastic calculus initiated by Doob. 
Ornstein and Uhlenbeck showed that in the short time limit the Gaussian distribution no longer held, and had to be replaced by a relation of the type $\langle x^2 \rangle \propto t^2$, thus leading to a proper definition of the velocity in this limit. 

During this walk through history, we did not witness any dialogue between mathematicians and physicists, for they worked on the same subject in parallel without interacting, expect for the small mention of Wiener's work by Uhlenbeck in which he denied its usefulness for physical purposes. Indeed, Wiener's theory did not significantly improve our understanding of physical phenomena related to Brownian motion, but instead contributed to the birth of the field of stochastic processes, which is now highly used by physicists. 

In the end, velocity is a concept whose existence and meaning depend on the model one considers. When talking about experiments, velocity may have the classic definition used in physics but cannot be measured, which explains the failure of early attempted measurements. When talking about Wiener process, velocity is not defined, and trajectories are described by continuous functions without tangent at any point, leading to the development of fractal theory. If we consider the Ornstein-Uhlenbeck process, velocity is well defined and linked to physical parameters, whereas it is not part of the Einstein-Smoluchowski theory.

\begin{appendices}
\section{Results overview}
\label{sec_comparatif}
\noindent
We intend in this appendix to give a one-page synthesis of the different theories examined in this article, for the reader to grab at once all the ideas and results. 
The first table compares Einstein's, Smoluchowski's and Langevin's theories as for their physical ingredients, mathematical contents and results. The second figure provides a guideline to Wiener's work, following the steps of his construction. The last table gathers the articles in which Ornstein's and Uhlenbeck's results are contained.

\begin{table}[!h]
	\centering
	{\tabulinesep=1.5mm
		\begin{tabu}{|c|c|c|c|c|}
			\hline
			\multicolumn{2}{|c|}{} & Einstein & Smoluchowski  & Langevin  \\
			\hline
			\multirow{4}{*}{\rotatebox[origin=c]{90}{Physics}} & Equipartition of energy  & Indirect & \checkmark & \checkmark \\
			\cline{2-5}
			& Stokes' law & \checkmark & \checkmark & \checkmark  \\
			\cline{2-5}
			& Osmotic pressure  & \checkmark & \xmark & \xmark \\
			\cline{2-5}
			& Treatment of collisions  & \xmark & \checkmark & Through $X$  \\
			\hline	
			\multirow{2}{*}{\rotatebox[origin=c]{90}{Maths}} & Stochasticity  & Distributions & Averages & Noise $X$ \\
			\cline{2-5}
			& Existence of velocity & \xmark & \xmark & \checkmark \\
			\hline
			\multirow{3}{*}{\rotatebox[origin=c]{90}{Results}} & $\lambda_x=f(t)$ & \checkmark & \checkmark & \checkmark \\			
			\cline{2-5}
			& $D=f(T,\mu,a)$ & \checkmark & \checkmark & \xmark \\
			\cline{2-5}
			& Diffusion equation & \checkmark  & \xmark & \xmark \\
			\hline
		\end{tabu}}
	\caption{Comparison of Einstein's, Smoluchowski's and Langevin' theories of Brownian motion.}
\end{table}
~\\
\begin{figure}[!h]
	\begin{center}
		\includegraphics[width=0.8\textwidth]{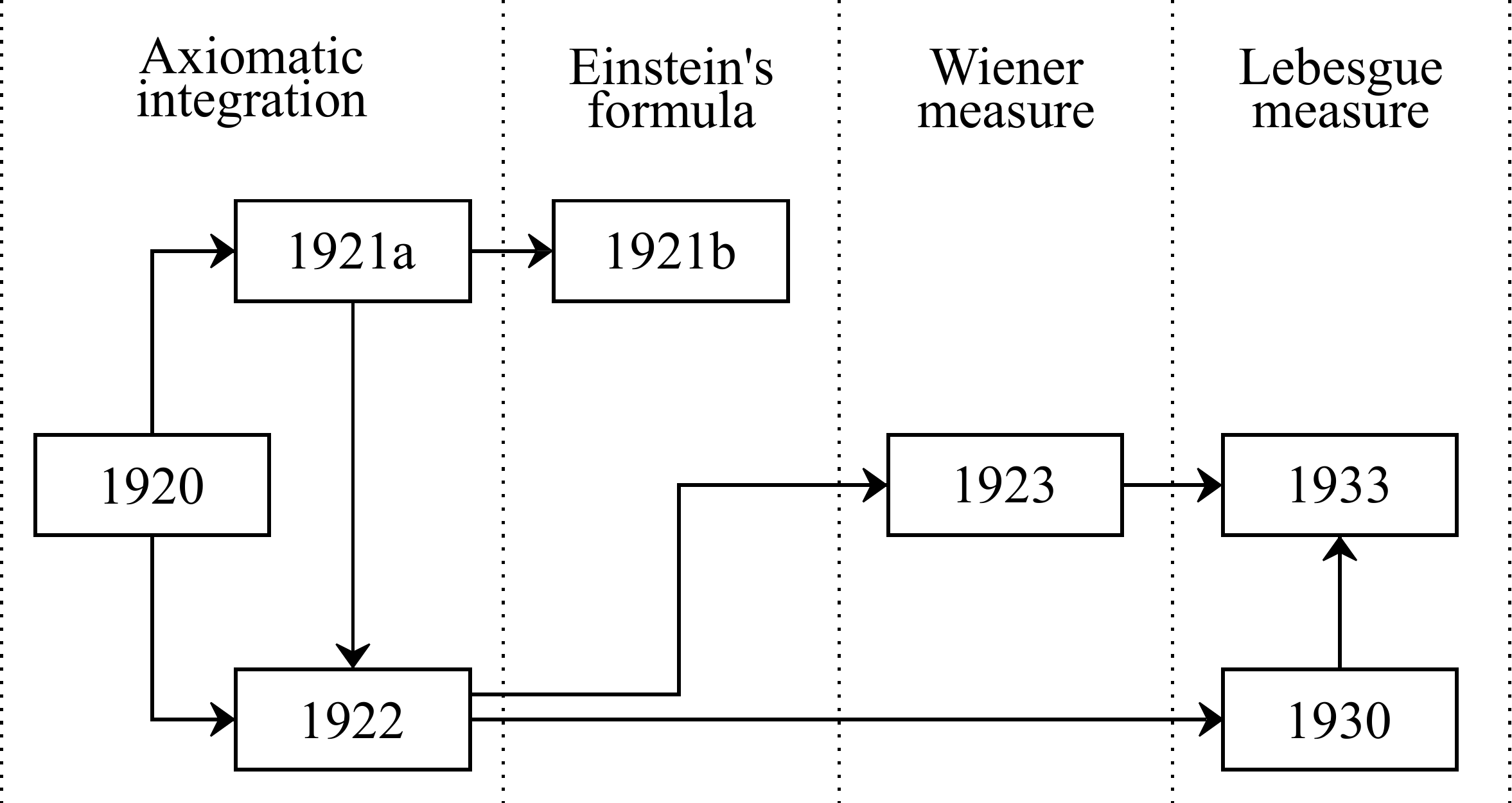}
		\caption{Guide to Wiener's work. The seven articles are split into the four stages of Wiener's construction, and the ideas continuity from an article to an other is symbolised by arrows.}
	\end{center}
\end{figure}
~
\begin{table}[!h]
	\centering
	{\tabulinesep=1.5mm
		\begin{tabu}{|c|c|c|c|c|c|}
			\hline
			& $\langle q \rangle$ & $\langle q^2 \rangle$ & $\langle q^k \rangle$, $k \geq 3$ & $f(q, q_0, t)$ & Fokker-Planck equation\\
			\hline
			$q=x$ & 1917 (O) & 1917 (O) & 1930 (O, U)& 1930 (O, U) & 1934 (O, W)\\
			\hline
			$q=v$ &  1917 (O) & 1917 (O) & 1930 (O, U) & 1930 (O, U) & 1917 (O) \\
			\hline
		\end{tabu}}
	\caption{Summary of Ornstein's and Uhlenbeck's main results. Each box contains the year and the authors (O : Ornstein, U : Uhlenbeck, W : van Wijk) of the article in which the result appearing in the first line is derived.}
\end{table}

\vfill

\end{appendices}

\newpage 

\printbibliography

\end{document}